\documentclass[preprint]{aastex61}
\pdfoutput=1 
\usepackage{amsmath,amstext}
\usepackage[T1]{fontenc}
\usepackage{apjfonts} 
\usepackage{graphicx}
\usepackage[flushleft]{threeparttable}
\usepackage{hyperref}

\newcommand{\changede}[1]{#1}
\newcommand{\changedf}[1]{#1}
\newcommand{\changedg}[1]{#1}
\newcommand{\changedh}[1]{#1}

\newcommand{\changedj}[1]{#1}
\newcommand{\changedk}[1]{#1}
\newcommand{\changedl}[1]{#1}
\newcommand{\changedm}[1]{#1}

\begin{document}

\title{A Binary Offset Effect in CCD Readout and its Impact on Astronomical
    Data}
    
\author{K. Boone}
\affiliation{Physics Division, Lawrence Berkeley National Laboratory, 1 Cyclotron Road, Berkeley, CA, 94720}
\affiliation{Department of Physics, University of California Berkeley, 366 LeConte Hall MC 7300, Berkeley, CA, 94720-7300}

\author{G. Aldering}
\affiliation{Physics Division, Lawrence Berkeley National Laboratory, 1 Cyclotron Road, Berkeley, CA, 94720}

\author{Y. Copin}
\affiliation{
    \changedj{
        Universit\'e de Lyon, F-69622, Lyon, France ; Universit\'e de Lyon 1, Villeurbanne ; CNRS/IN2P3, Institut de Physique Nucl\'eaire de Lyon
    }
}

\author{S. Dixon}
\affiliation{Physics Division, Lawrence Berkeley National Laboratory, 1 Cyclotron Road, Berkeley, CA, 94720}
\affiliation{Department of Physics, University of California Berkeley, 366 LeConte Hall MC 7300, Berkeley, CA, 94720-7300}

\author{R. S. Domagalski}
\affiliation{Physics Division, Lawrence Berkeley National Laboratory, 1 Cyclotron Road, Berkeley, CA, 94720}
\affiliation{Department of \changedj{Astronomy}, University of Toronto, \changedj{50} St. George St., Toronto, ON, M5S 3H4, Canada}
\affiliation{Dunlap Institute for Astronomy \& Astrophysics, 50 St. George St., Toronto, ON, M5S 3H4, Canada}

\author{E. Gangler}
\affiliation{Laboratoire de Physique Corpusculaire de Clermont-Ferrand, F-63171 Aubiere Cedex, France}

\author{E. Pecontal}
\affiliation{Centre de Recherche Astronomique de Lyon, Universit\'e Lyon 1, 9 Avenue Charles Andr\'e, F-69561 Saint Genis Laval Cedex, France}

\author{S. Perlmutter}
\affiliation{Physics Division, Lawrence Berkeley National Laboratory, 1 Cyclotron Road, Berkeley, CA, 94720}
\affiliation{Department of Physics, University of California Berkeley, 366 LeConte Hall MC 7300, Berkeley, CA, 94720-7300}



\begin{abstract}

    We have discovered an anomalous behavior of CCD readout electronics that
    affects their use in many astronomical applications. An offset in the
    \changedh{digitization of the CCD output voltage that depends} on the
    binary encoding of one pixel is added to pixels that are read out one, two
    and/or three pixels later. One result of this effect is \changedh{the
        introduction of a differential offset in the background when comparing
        regions with and without flux from science targets.} Conventional data
    reduction methods do not correct for this offset. We find this effect in 16
    of 22 instruments investigated, \changedj{covering} a variety of telescopes
    and many different front-end electronics systems. The affected instruments
    include LRIS and DEIMOS on the Keck telescopes, WFC3-UVIS and STIS on HST,
    MegaCam on CFHT, SNIFS on the UH88 telescope, GMOS on the Gemini
    telescopes, HSC on Subaru, and FORS on VLT. The amplitude of the introduced
    offset is up to 4.5~ADU per pixel, and it is not directly proportional to
    the measured ADU level. We have developed a model that can be used to
    detect this "binary offset effect" in data and correct for it.
    Understanding how data are affected and applying a correction for the
    effect is essential for \changedf{precise astronomical measurements.}

\end{abstract}


\section{Introduction}

Charge coupled devices (CCDs) have been the dominant astronomical detector for
the past three decades.
Ideally, light at the focal plane of the telescope is detected on a grid of
pixels which each provide independent information about the incident
light at
their location. In practice, astronomical CCDs \changedj{and the associated
    instruments} are subject to many
anomalies that introduce spurious signals, correlations between pixels
or deviation from linear behavior. These anomalies
will lead to errors in the derived scientific results if they are not understood
and accounted for. There are several locations in the instrument
where such anomalies can be introduced.

Many of the anomalies are related to imperfections in the production of the
sensor. These include variations in the pixel areas, fringing due to variations
in thickness of the CCD, "tree ring" patterns due to impurities in the
production of the silicon wafers, manufacturing defects in the CCD electronics,
and edge effects due to interactions with other components or guard rings. See
\citet{janesick01} or \citet{stubbs14} for in-depth discussions of these
effects. Localized contamination of the silicon can also lead to many
undesirable effects, including hot pixels with high dark current and traps that
interfere with the charge transfer of pixels read out later in the same column.
Large traps can produce dead or hot columns, while smaller traps lead to charge
transfer inefficiency (CTI). \changedj{\citet{baggett12} discusses localized
    contamination in the HST WFC3 UVIS detectors and studies how it evolves
    with time.}

\changedj{
    Several anomalies arise due to normal instrument operations even with a
    defect-free CCD. For very bright sources, the CCD can saturate, leading to
    effects such as blooming where charge spills into neighboring pixels. In
    general, the presence of charge on the CCD distorts the nearby electric
    field. This distortion manifests itself as effects like the
    "brighter-fatter effect" where the widths of point-spread functions vary by
    up to 2\% for bright and faint objects \citep{antilogus04}. It is also
    possible to accumulate charge on the CCD from sources other than the
    desired science target, such as cosmic rays which deposit charge as they
    pass through the CCD \citep{janesick01}, or optical reflections ("ghosts")
    that appear when light reflects in unintended ways off of external optical
    components such as filters \citep{brown04} or within the CCD itself.
}

Readout electronics can introduce another set of anomalies. Many readout
systems are susceptible to pickup \changedh{that} appears as periodic oscillations in the
measured values of pixels that are observed sequentially. This can introduce a
herring-bone pattern \changedh{across the CCD image} \citep{jansen03}. Other potential artifacts of
the readout electronics include undershooting after reading a bright pixel
\citep{caldwell10}, crosstalk between the
\changedh{readouts from} different amplifiers \citep{baggett04}, and biased/sticky bits in analog-to-digital
converters \citep{robberto05}. Understanding how all of these effects
interact with scientific data is essential for proper calibration of the
science output of an instrument. Techniques have been developed to correct for
most of these effects, and data reduction pipelines typically endeavor to treat
the ones that have a sizeable effect on the data taken by their targeted instrument.

In this paper, we report a newly discovered CCD electronic chain artifact which
originates in the readout electronics of many commonly-used CCD electronics
systems. We find that there is crosstalk between the binary-coded output of the
analog-to-digital converter (ADC) and \changedk{subsequently read-out pixels}. An
offset is introduced into the \changedk{subsequently read-out pixels} which is roughly proportional to
the number of "1" bits in the binary encoding of a driver pixel. This effect is
present in a wide variety of currently-used instruments with different
electronics configurations. We have characterized the effect and modeled it
with high accuracy in the SuperNova Integral Field Spectrograph
\changedf{\citep[SNIFS;][]{lantz04}} instrument used by the Nearby Supernova
Factory collaboration \changedf{\citep[SNfactory;][]{aldering02}} on the UH88
telescope. We call this CCD-electronics-chain artifact the "binary offset
effect".

We proceed as follows. In Section~\ref{sec:identifying} we discuss how to
identify this effect in CCD data. In Section~\ref{sec:model}, we build a model
of the effect which can predict the size of the introduced offsets given the adjacent
pixel values, and \changedf{we provide an example of data corrected with
this model.} In Section~\ref{sec:impact},
we then discuss how scientific results \changedh{can be} impacted by this effect if it is not
corrected for.

\section{Identifying the Binary Offset Effect}
\label{sec:identifying}

\subsection{Evidence of the Binary Offset Effect in SNIFS Data}
\label{sec:evidence}

The binary offset effect was first observed in data taken with SNIFS for
\changedf{the} SNfactory \citep{lantz04, aldering02}. SNIFS contains two lenslet integral
field unit (IFU) spectrographs which produce spectra over a 15$\times$15 grid of
spatial elements \changedf{(spaxels)}. The spectrographs \changedf{simultaneously} cover wavelength ranges of 3200-5200~\AA~and
5100-10000~\AA~for the blue and red channels respectively. Each
spectrograph uses a CCD composed of 2048$\times$4096 15 micron pixels. The blue
channel uses a thinned E2V model 44-82 CCD while the red channel uses a thinned and deep
depleted E2V CCD44-82-0-A72 CCD. These CCDs are of the highest scientific
grade. Each CCD is read out by two independent amplifiers using an Astronomical
Research Cameras (ARC) Generation II video board \citep{leach98}.

\begin{figure}
\centering
\includegraphics[scale=0.75]{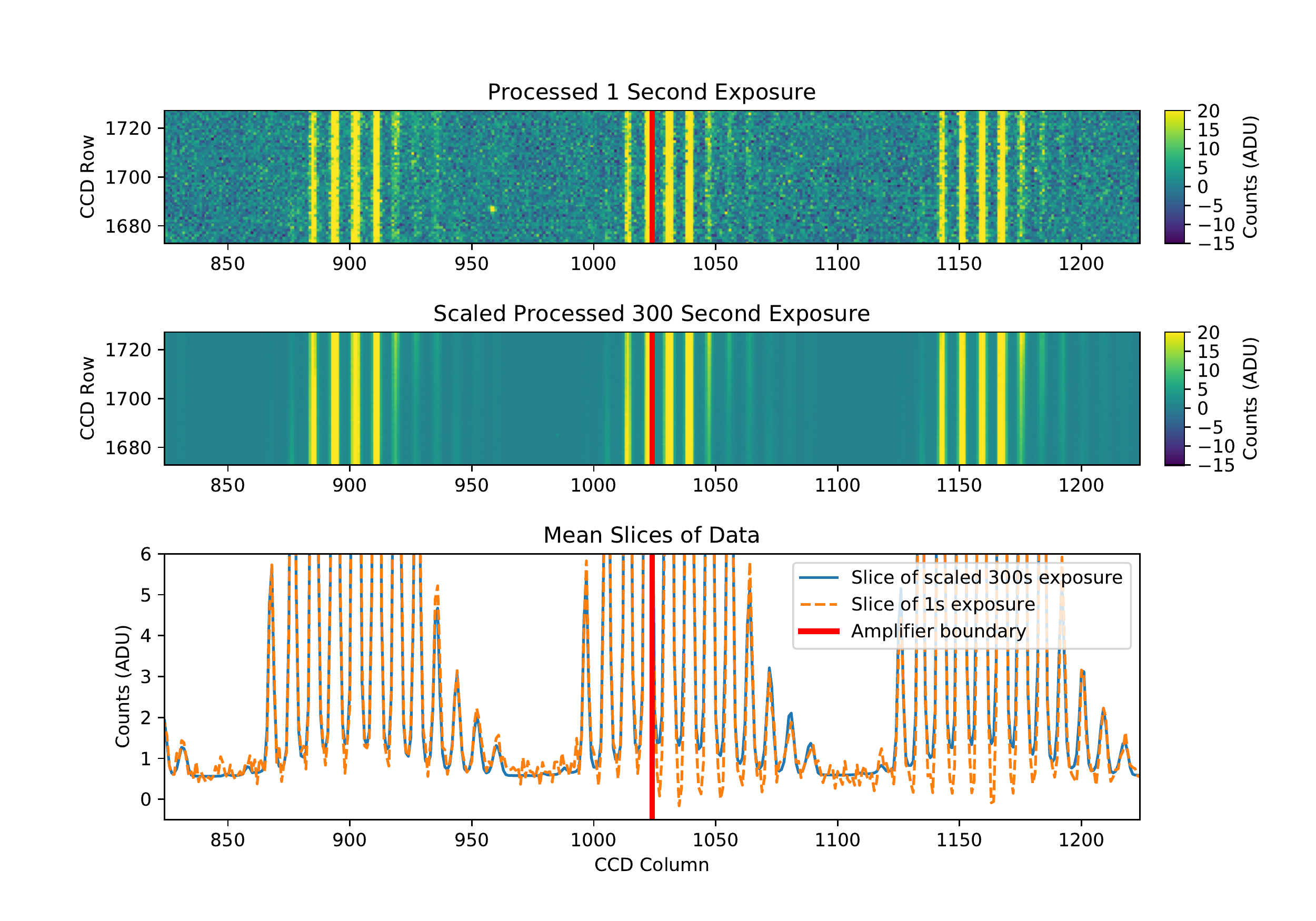}

\caption{
    \changedj{
        Comparison of 1 and 300 second dome flat exposures with the SNIFS
        instrument. Top panel: 1 second exposure. Middle panel: 300 second
        exposure. The vertical lines (which saturate the scale) are the traces
        of spectra for each of the individual spaxels. The amplifier boundary
        on the CCD is indicated with a vertical red line.  Bottom panel: slices
        through the previous images showing the mean values of the pixels in
        each CCD column with CCD \changedk{row} values between 1400 and 2000. The 300 second
        exposure has been scaled by its exposure time to match the 1 second
        exposure. On the right amplifier, there is a deficit in the measured
        counts between each spaxel for the 1 second exposure relative to the
        300 second exposure, although the background regions are in agreement.
        These large deficits are not present on the left amplifier.
    }
}
\label{fig:uncorrected2d}
\end{figure}

\changedj{
    The binary offset effect can be easily seen by looking at images where
    there is an accurate model of the light on the CCD, and by probing the
    residuals after this model is subtracted from the data. Here, we examine a
    set of 1~second dome flat exposures taken with SNIFS. We take a 300~second
    exposure in the same configuration, and we treat this exposure
    as a model of the true light on the CCD because this exposure has high
    count levels compared to the size of the effects that we are looking into.
    An example of these dome flat CCD exposures is shown in
    Figure~\ref{fig:uncorrected2d}, where the spectral traces due to the IFU
    reformatting of the dome flat are all visible. We systematically work with
    the output of the CCD in analog-to-digital units (ADU) throughout this
    paper. The top two panels \changedk{of} Figure~\ref{fig:uncorrected2d} show examples of
    the dome flat exposures after applying \changedl{an overscan subtraction, a bias correction,
    and a dark correction}. The bottom panel of this figure shows slices
    through both one of the 1~second dome flat exposures and the 300~second
    exposure scaled to match the exposure time of the 1~second exposure. On the
    right amplifier, there is a count deficit of 1-2~ADU in between each of the
    spaxels on the 1~second exposure relative to the 300~second one. This same
    deficit is not seen on the left amplifier.
}

\begin{figure}
\centering
\includegraphics[scale=0.8]{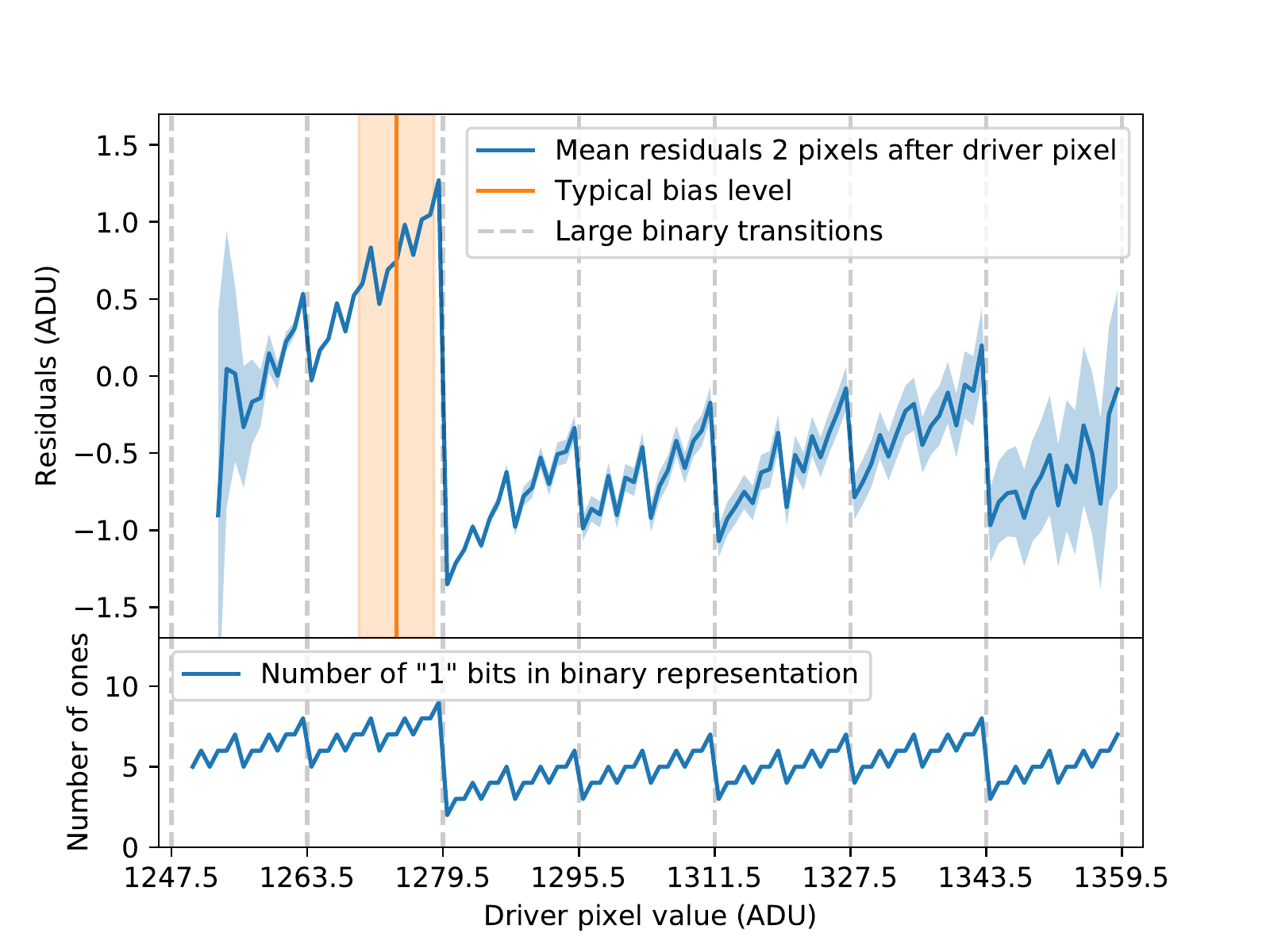}

\caption{Example of the binary offset effect on the SNIFS blue channel right
    amplifier. \changedj{See text for details on how this plot was produced.
        Top panel: mean residuals as a function of the raw driver
        pixel value read out 2 pixels earlier along with an uncertainty band.
        Bottom panel: number of "1" bits in the binary representation of the driver
        pixel value.} We label transitions where more than 5 bits flip with a
    vertical dashed line. All of the features in the bottom panel, including
    the large transitions, are visible in the residual data. The mean
    background level and one standard deviation of \changedf{read noise} are
    illustrated with a vertical orange line and surrounding band.}

\label{fig:binary_offset_effect}
\end{figure}

\changedj{To probe the cause this deficit, we} subtract a scaled version of the
300~second exposure (which we use here as our \changedk{reference} model) from each of the
1~second exposures to obtain residual images. For each 1~second exposure, we
then compare the residual values of each pixel after the model was subtracted
to the raw value (number of ADU) measured in a pixel that was read out 2~pixels
earlier. We call the pixel read out 2~pixels earlier the "driver pixel". We
take the mean of all residuals that have the same driver pixel value, and we
plot the mean residual as a function of the driver pixel value. The results of
this procedure are shown in Figure~\ref{fig:binary_offset_effect}.

The resulting plot gives a very jagged function: a single ADU difference in the
driver pixel value can correspond to \changedj{a difference} of up to 2.5~ADU
in the mean of the residuals of the pixel read out 2 pixels afterwards. These
\changedj{differences} are highly statistically significant; the measurement
uncertainties are less than 0.05~ADU between driver pixel values of 1260 and
1290. A careful analysis of these large \changedj{differences} reveals that
they map out the number of "1" bits in the binary representation of the driver
pixel value. For example, Figure~\ref{fig:binary_offset_effect} shows that
there is a difference of 2.5~ADU when transitioning from a driver pixel value
of 1279~ADU to a driver pixel value of 1280~ADU. The binary representations of
these two numbers are 100~1111~1111 and 101~0000~0000 respectively. Other large
\changedj{differences} occur between 1311~ADU and 1312~ADU (101~0001~1111 and
101~0010~0000) and between 1343~ADU and 1344~ADU (101~0011~1111 and
101~0100~0000). \changedj{The lower panel of
    Figure~\ref{fig:binary_offset_effect} shows the plot of the number of "1"
    bits in the binary representation of the driver pixel for comparison. All
    of the features in this plot are seen in the residual data, from the large
    binary transitions down to odd-even effects.} 

\begin{figure}
\centering
\includegraphics[scale=0.8]{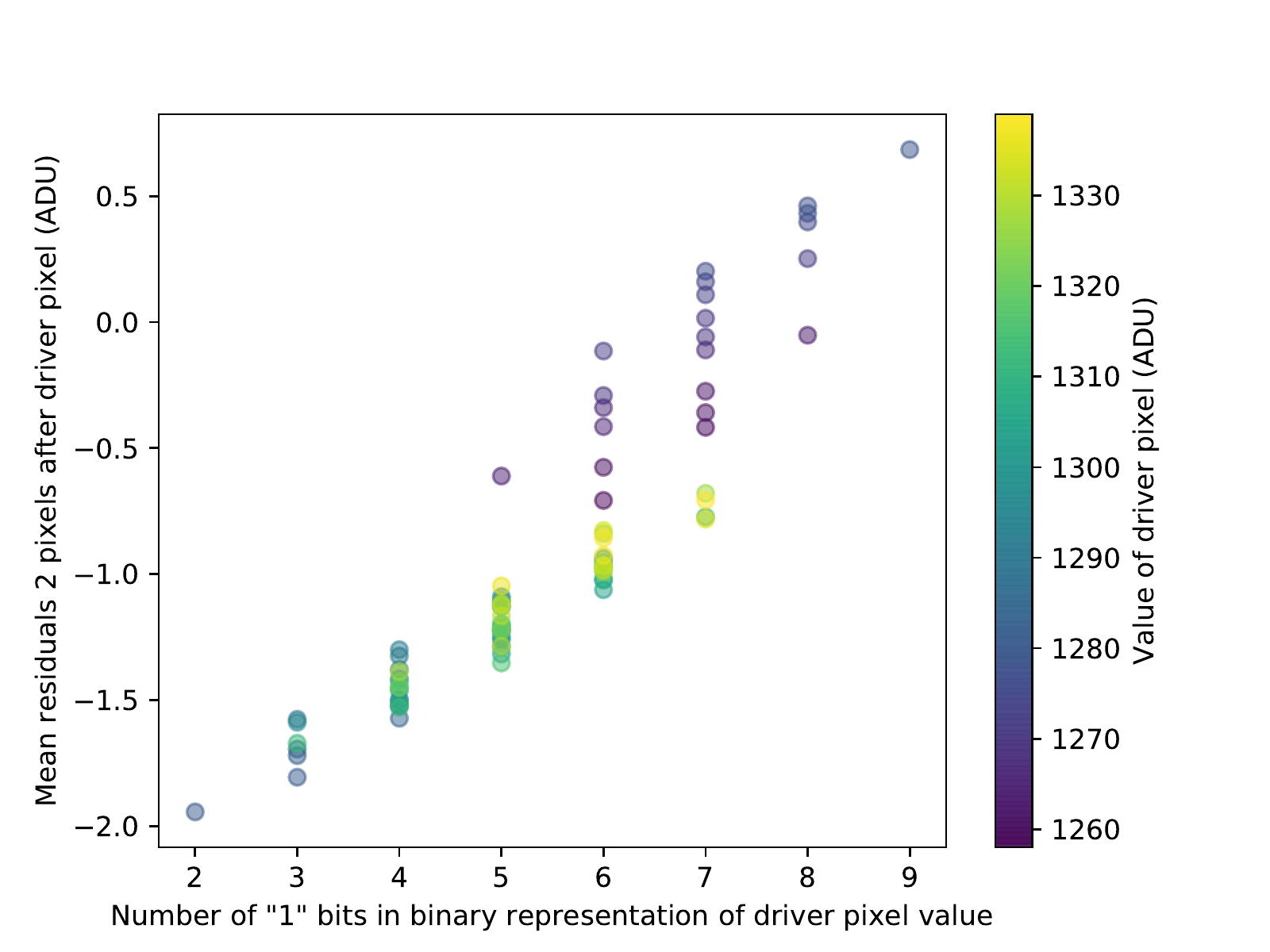}

\caption{Example of the binary offset effect on the SNIFS blue channel right amplifier. This
    plot shows the same data as in Figure~\ref{fig:binary_offset_effect} with a
    cut requiring that the uncertainty on the mean of the residuals be less the 0.2~ADU.
    Instead of plotting the mean of the residuals directly against the driver pixel values as in Figure~\ref{fig:binary_offset_effect},
    we sum the number of "1" bits
    in the binary representation of each driver pixel value in ADU, and we plot the mean of the residuals
    against this number. There is a clear
    linear trend in the mean of the residuals with the number of "1" bits in the binary
    representation, at all scales. It is also apparent that there are some
    additional effects, such as an offset in the zeropoint of the linear relation
    near a driver pixel value of 1280~ADU.}

\label{fig:residuals_vs_binary}
\end{figure}

Figure~\ref{fig:residuals_vs_binary} shows the same residuals from
Figure~\ref{fig:binary_offset_effect} plotted directly against the number of "1" bits in
the binary representation of the driver pixel value. There is a linear trend
between the mean of the residuals and the number of "1" bits in the binary representation of
the driver pixel value. This is consistent with this effect being mostly related to
the total number of "1" bits rather than it being due to some issue
with the uppermost bits. There is an offset in the zeropoint of this linear
relation for values below 1280~ADU indicating that a linear model with the
number of "1" bits is not a complete description of the effect. In Section~\ref{sec:model}
we will discuss how to properly model this effect.

\subsection{Implications of the Binary Offset Effect in SNIFS Data}
\label{sec:implications_snifs}

\begin{figure}
\centering
\includegraphics[scale=0.85]{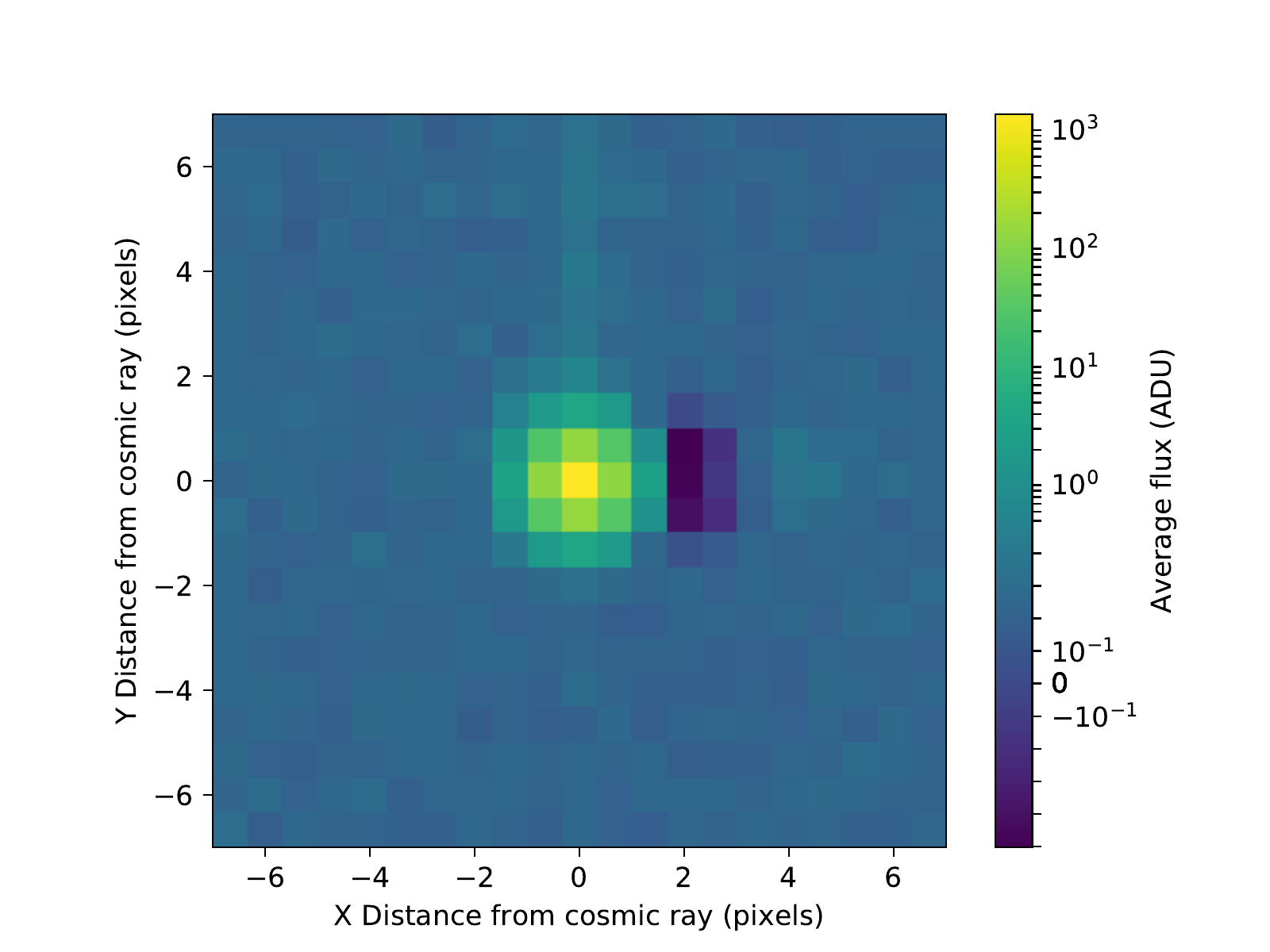}

\caption{Stacked cosmic ray frames from the right amplifier of the
    SNIFS blue channel. This image was
    generated by taking a series of dark exposures on a night, identifying all
    face-on cosmic rays in the image, and \changede{taking the clipped mean of
    all pixels around cosmic rays}. The color scale on
this plot changes from linear to logarithmic at \changedj{0.5~ADU} \changedk{in order} to capture the
    full dynamic range of the image. \changede{The serial readout reads pixels from left to right in this image.}
    There is a visible deficit of $\sim$0.5~ADU 2-3 pixels following the cosmic ray.}

\label{fig:deficit}
\end{figure}

\changedj{
    The binary offset effect introduces a highly non-linear signal into images
    taken with the CCD. The main consequence of the binary offset effect is
    that an offset can be introduced to \changedk{CCD data} that only appears in regions
    where sufficient flux from science targets is present on the CCD. To show
    how this appears in data, we examine stacks of face-on cosmic rays in
    dark exposures taken with the SNIFS instrument. An example of such a stack
    is shown in Figure~\ref{fig:deficit}. These images were originally taken in
    effort to measure the in-situ CTI of the SNIFS instrument \citep{dixon16},
    and a small CTI tail can be seen on the upper part of the image. We also
    find a deficit in the measured count levels of pixels read out 2-3 pixels
    after a cosmic ray compared to the background level. We find that the size
    of the deficit varies from night to night, from no effect up to around
    1~ADU per pixel. The size of the deficit is not the same on each amplifier.
    This deficit is the result of the binary offset effect, and it can be
    thought of as an offset that is added to the data when the count
    levels of previous pixels cross a specific threshold. The deficit has a
    roughly fixed count value that does not scale with the amount of flux in
    the preceding pixels, so it does not behave as a simple modification to the
    PSF. The binary offset effect hence effectively introduces a deficit
    wherever there is sufficient flux from science targets on the CCD, leading
    to a local offset in the measured count values for regions on the CCD with
    flux from science targets compared to those without any flux.
}

\changedj{
    An explanation of how the binary offset effect can cause such a deficit can
    be seen from Figure~\ref{fig:binary_offset_effect}. The raw background
    level (before overscan subtraction) on the dome flat images used to
    generate this figure is around 1274~ADU with 4.5~ADU (or 3~electrons) of
    read noise per unstacked pixel. This background level is illustrated with
    an orange line and surrounding band on the figure. Most background pixels
    are therefore below the threshold of 1280~ADU, where there is a large
    change in the binary representation of the driver pixel value. When there
    is at least a small amount of signal on the CCD, the pixel values are
    brought above the 1280~ADU threshold and a $\sim$2~ADU offset is introduced
    into subsequently read pixels compared to the background regions. The
    consequences of the binary offset effect for a science image can be seen in
    the lower panel of Figure~\ref{fig:uncorrected2d}: for the right amplifier,
    a deficit appears in the 1 second dome flat exposure compared to the 300
    second exposure wherever the average count levels are above a few counts.
    On the left amplifier, the bias level is at 1285~ADU, and there are no
    major binary transitions in the nearby higher count levels. The binary
    offset effect is still present, but introduced offsets are similar for all
    pixels so no major difference is observed between background and science
    regions.
}

\begin{figure}
\centering
\includegraphics[scale=0.8]{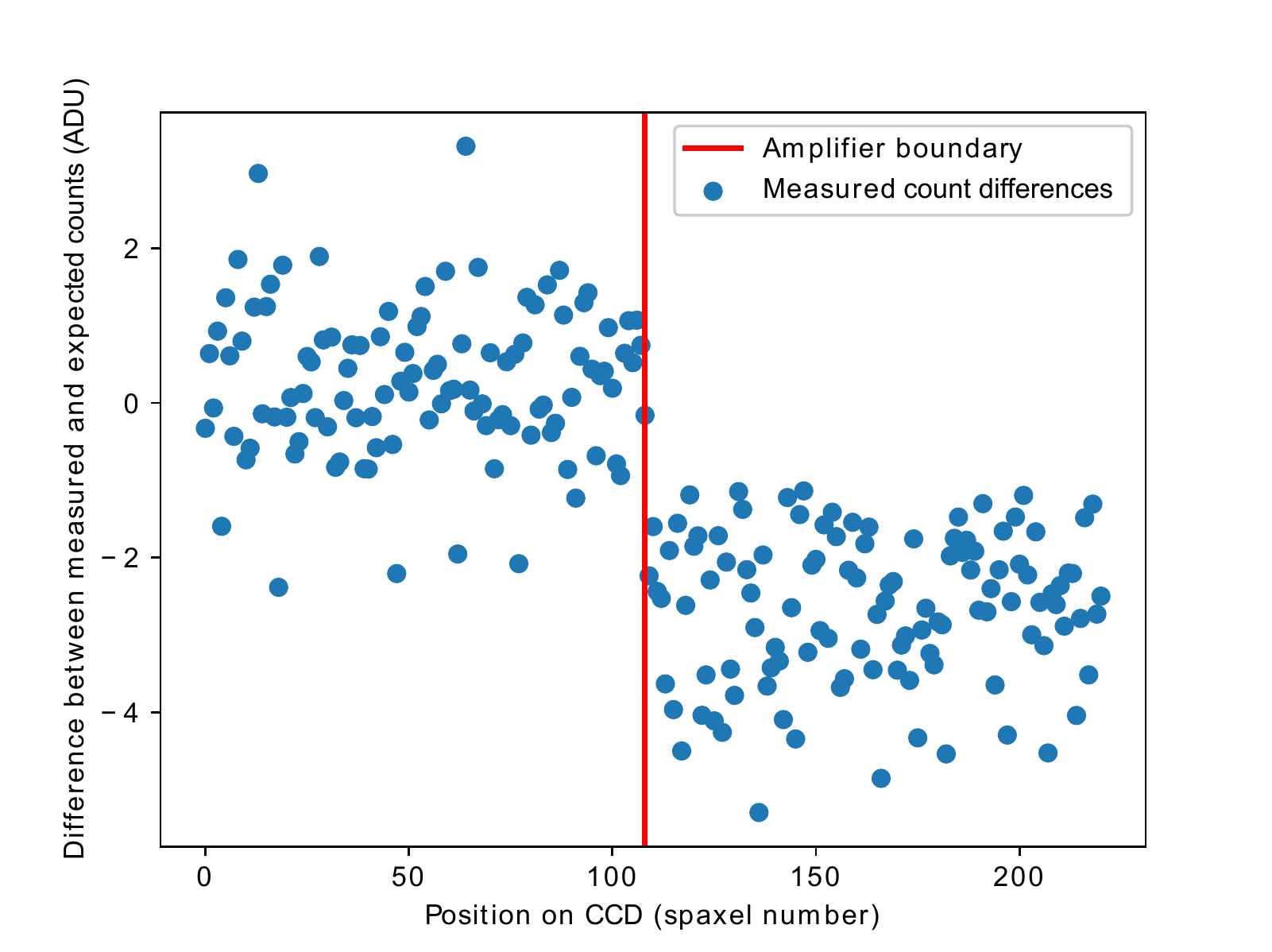}

\caption{Exposure-time normalized difference between extracted 1 second and
    300 second dome flats from the SNIFS \changede{blue} channel\changede{, illustrating the "blue step"}. Each point corresponds to the average flux of a spaxel in the 4000 to 4500 \AA~region. We
    indicate the boundary where the readout switches from the left to the right
    amplifier with a \changedh{vertical} red line. We find a 2.2~ADU difference in the measured
    fluxes between the two amplifiers which corresponds to a 4.2\% difference
    in the measured flux \changedh{for this example}.}

\label{fig:rawfluxratio}
\end{figure}

\changedj{
    When extracting the spectra from images affected by the binary offset
    effect, the introduced offsets will propagate to the extracted spectra. The
    two amplifiers will, in general, have different bias levels, so they will
    be affected in different ways by the binary offset effect. Hence,
    systematic differences will be observed in spectra extracted from each of the
    two amplifiers. The SNfactory collaboration discovered these systematic
    differences before the cause was known, and dubbed this effect the "blue step"
    as it was primarily noticed at the bluest wavelengths where the expected
    fluxes of SNIFS's primary objects of interest (Type~Ia supernovae) can be
    faint. To illustrate the blue step, we extract spectra for each of the IFU
    spaxels in one of the 1~second dome flat exposures using the standard
    SNfactory pipeline \citep{aldering06, scalzo10}, and we calculate the
    average flux in the 4000 to 4500~\AA~region. The results of this procedure
    are shown in Figure~\ref{fig:rawfluxratio}. We find a 2.2~ADU difference in the
    measured counts between the two amplifiers, corresponding to a 4.2\%
    difference in the measured fluxes between the two amplifiers for this
    exposure. The observed difference is well-modeled by a flat difference in the
    background level between the different amplifiers, as expected for a
    signal introduced in the previously described manner.
    The average offset introduced by the binary
    offset is different in regions on the CCD with and without flux present,
    meaning that there is a \changedk{effectively a} pathological local
    background offset that only appears where there is flux on the CCD. 
    \changedk{Conventional background subtraction routines cannot identify or correct
    for the difference in background levels introduced by the binary offset
    effect because these methods measure the background level in locations on
    the CCD where there is no flux present.}
}

\subsection{Cause of the Binary Offset Effect}

The discussion up to this point has \changedj{shown that an offset} is
introduced into pixels that are read out two pixels after the driver pixel.
\changedj{We find a similar offset three} pixels after the driver pixel for
each CCD amplifier on both the blue and red SNIFS channels. There is no effect
1 pixel after or 4 or more pixels after, nor is there an offset for pixels
before the driver pixel. The ARC video board used by SNIFS processes pixels one
at a time, so the timeframe of 2-3 pixels between the driver and affected
pixels implies that there must be some feedback in the electronics chain from
somewhere after the ADC conversion. We varied the gain of the SNIFS front-end
electronics by a factor of 10, and we find that the amplitude of the effect is
constant in ADU. This implies that the introduction of the feedback must occur
past the gain electronics in the CCD electronics chain. We also notice that the
binary offset effect shows up across amplifiers: a driver pixel will introduce
an offset in the pixels that are read out 3 pixels later in the other amplifier
with a similar amplitude to the offset introduced on the driver pixel's
amplifier.

Combining all of these findings, we propose that the binary offset effect is
being caused by feedback into the reference voltage of the ADC. The ADC
\changedj{is} highly sensitive to small changes in its reference voltages: one
ADU for a 16-bit ADC corresponds to a change of only 0.0015\%. When earlier
pixels are read out, the ADC outputs their binary representations and stores
them temporarily. We propose that these charges (representing the ones in the
binary code) then introduce a slight offset into the reference voltage when
they are released and the next pixel is read out.

\subsection{The Binary Offset Effect in Other Instruments}

We wrote a program that can identify the binary offset effect in CCD data from
any telescope by looking for the characteristic saw-tooth shape in the mean of
the residuals as a function of the driver pixel value seen in
Figure~\ref{fig:binary_offset_effect}. We find that the effect is present in
most astronomical instruments that are currently in use across a variety of
different readout systems. As we are not able to build a model using different
length exposures of our own on all of these instruments, we use bias or dark
images to probe the effect\changedj{, and we assume} that the pixel values in
these images can be modeled by a smooth function. We fit for such a model using
\changedk{the background determined by} \texttt{sep} \citep{barbary16, bertin96}
with $64 \times 64$ pixel boxes. We
subtract this model from the original images to obtain residual images. With
dark and bias images, we are limited to probing the effect over a smaller
baseline of driver values (typically around 10~ADU). We are, however, able to
achieve very high signal-to-noise by averaging the residuals over the full
image and by examining multiple images. \changedj{We estimate the amplitude of
    the binary offset effect by measuring the largest difference in residuals
    between two adjacent driver pixel values. For the SNIFS images shown in
    Figure~\ref{fig:binary_offset_effect}, this corresponds to the size of the
    transition from 1279 to 1280. We note that the measurement uncertainties on
    the stacked residuals are around 0.01 ADU for most instruments, so this
    method is sensitive to effects larger than $\sim$0.05~ADU.} The results
are summarized in Table~\ref{tab:offset_size}.

\begin{figure}
    \centering
    \setlength{\extrarowheight}{-20pt}
    \changede{
    \begin{tabular}[b]{ll}
        \includegraphics[width=0.45\textwidth]{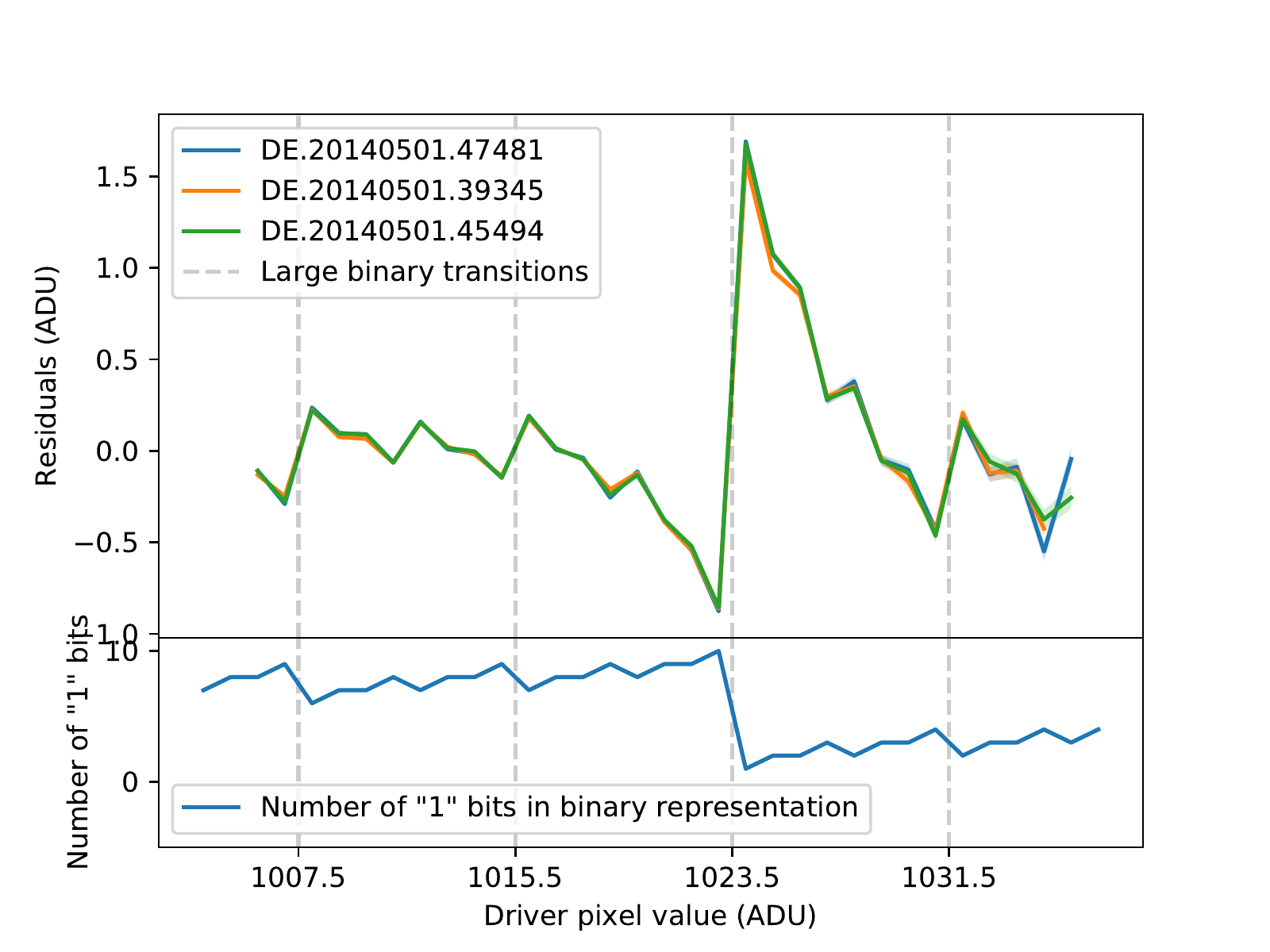} &
        \includegraphics[width=0.45\textwidth]{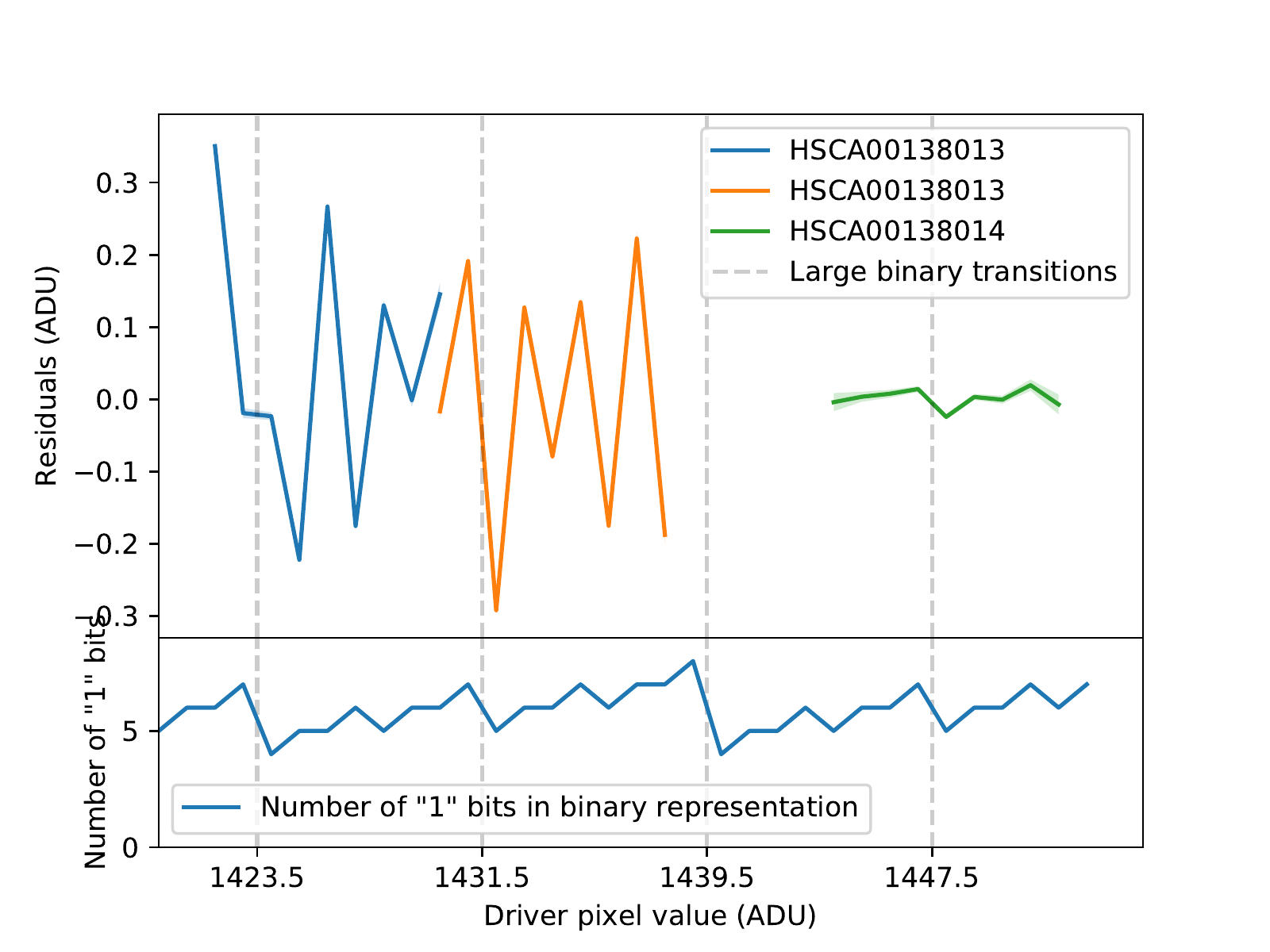}
        \\
        \small (a) DEIMOS (Keck), 2 pixels after the driver pixel. &
        \small (b) Hyper Suprime-Cam (Subaru), 1 pixel after \\
        The median measurement uncertainty for the &
        the driver pixel. The median measurement \\
        residuals is 0.012~ADU. &
        uncertainty for the residuals is 0.006~ADU. \\
        \includegraphics[width=0.45\textwidth]{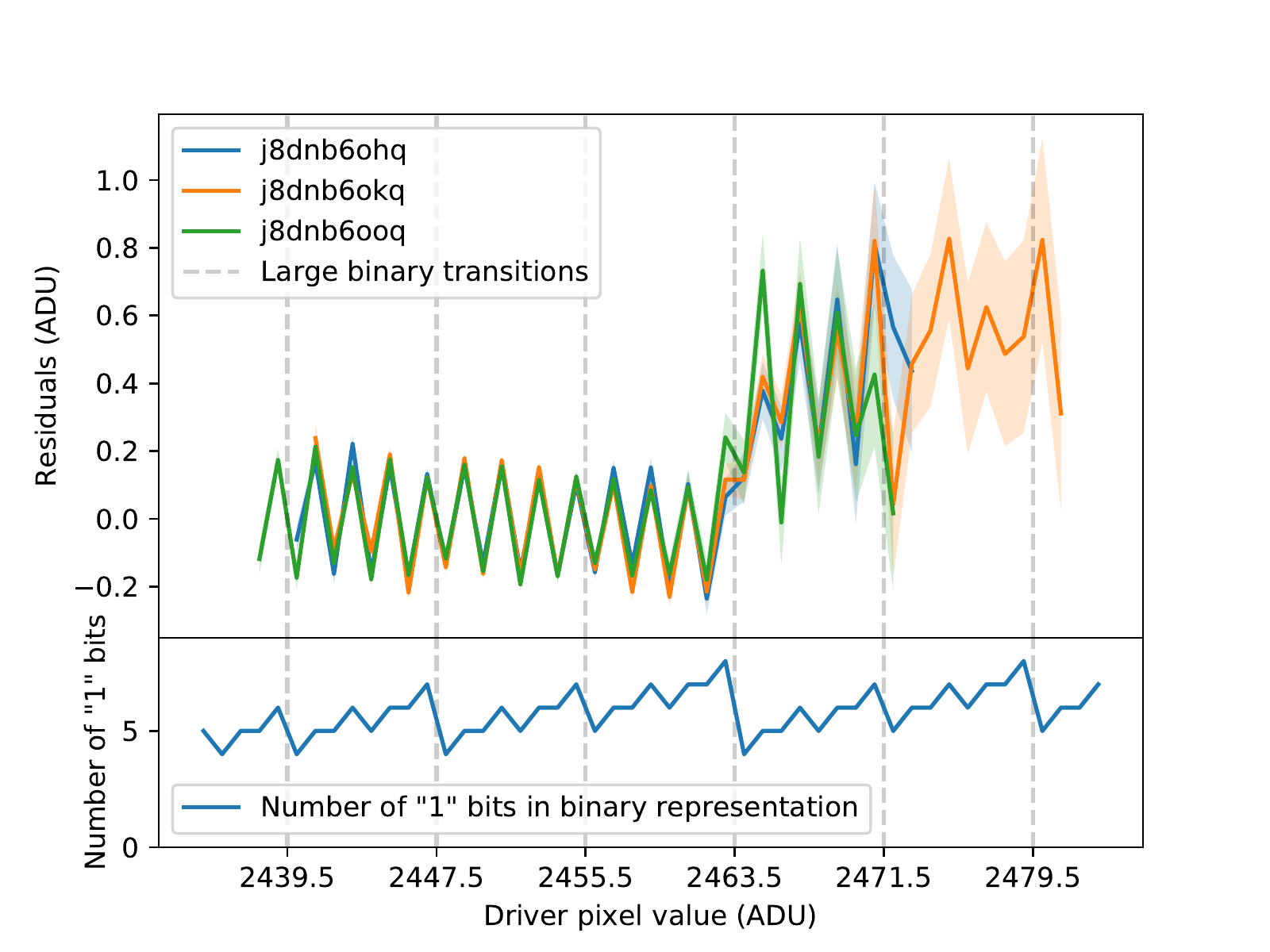} &
        \includegraphics[width=0.45\textwidth]{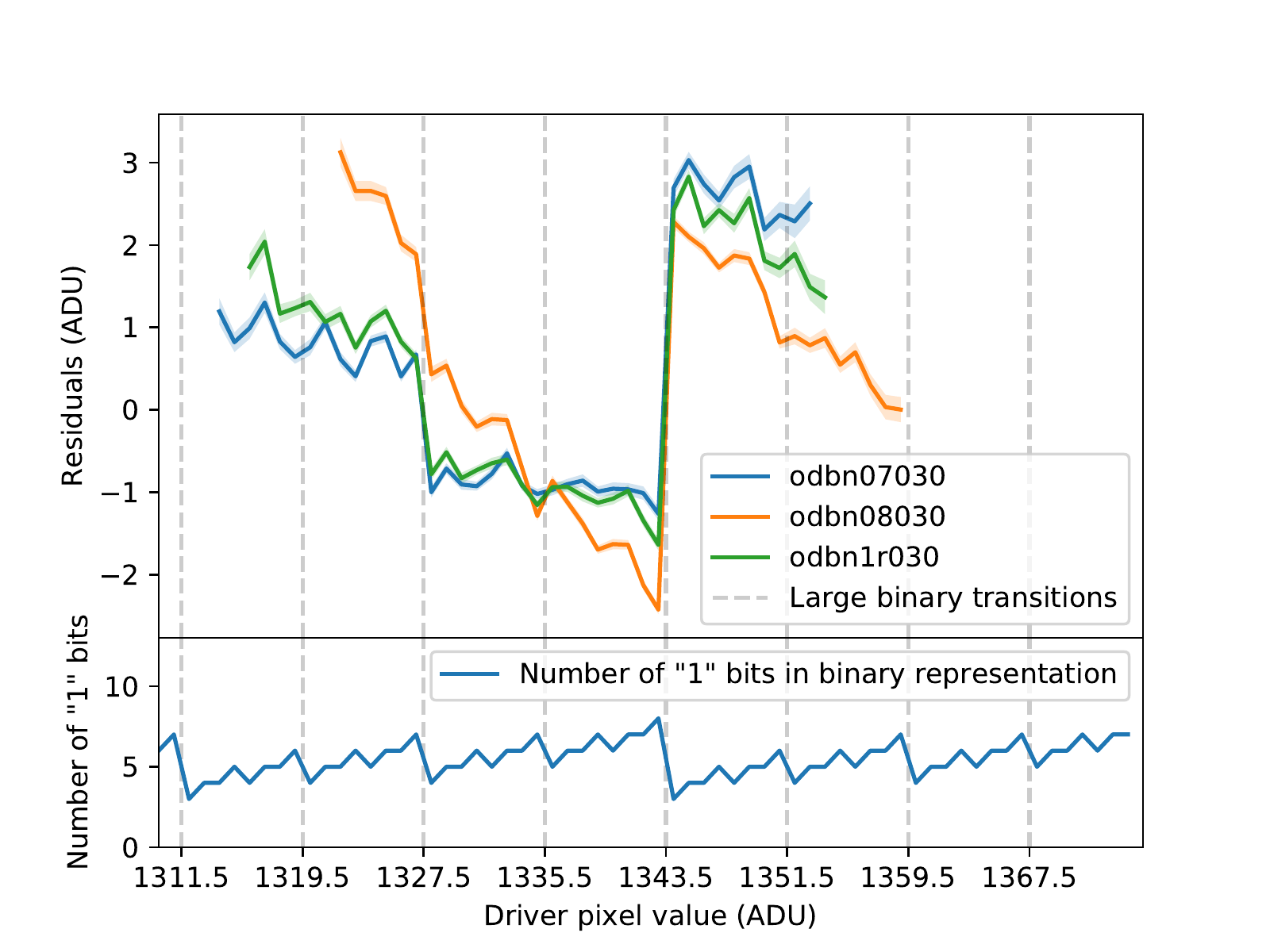}
        \\
        \small (c) ACS (HST), 1 pixel after the driver pixel. &
        \small (d) STIS (HST), 1 pixel after the driver pixel. \\
        The median measurement uncertainty for the &
        The median measurement uncertainty for the \\
        residuals is 0.030~ADU. &
        residuals is 0.077~ADU. \\
    \end{tabular}
    }

    \caption{Examples of the binary offset effect in other instruments. For
        half of the instruments that were investigated, such as DEIMOS on Keck,
        we see an effect that has similar properties to the effect on SNIFS.
        For other instruments, such as \changedj{Hyper Suprime-Cam on Subaru,
            ACS on HST, or STIS on HST,} we find evidence of offsets related
        \changedh{to} the binary representation, but the offsets have a
        different dependence on the binary encoding. All plots are shown with
        their measurement uncertainties as a shaded contour\changede{, and the
            median measurement uncertainty is \changedf{given} in the
            \changedg{subpanel} caption. These measurement uncertainties are
            very small compared to the size of the observed effects, and all of
            these observed effects are highly statistically significant.} The
        specific \changede{observations} used to generate each plot are
        \changedh{listed} in the figure legends.}

    \label{fig:other_binary_offset}
\end{figure}

We are able to detect the binary offset effect in 16 of 22 instruments that
were investigated. The amplitude of the offset varies significantly between the
different instruments. We provide a \texttt{Jupyter} notebook
\changedj{\citep{kluyver16}} that contains code to \changedj{probe} this effect
for all of the instruments listed in Table~\ref{tab:offset_size}. This
\texttt{Jupyter} notebook is available on \changedm{GitHub at 
\url{https://github.com/snfactory/binaryoffset}.} Plots of the binary offset
effect are shown for a representative sample of the different instruments in
Figure~\ref{fig:other_binary_offset}.

\begin{table}
\tiny
\begin{threeparttable}

    \caption{Size of the binary offset effect in different instruments. See
        text \changedj{and accompanying \texttt{Jupyter notebook}} for
  details on how these measurements were made. We have attempted to determine the
  CCD \changedh{electronics} front end and ADC wherever possible for comparison purposes.}

\label{tab:offset_size}
\begin{tabular}{|c|c|c|c|c|c|c|c|}
    \hline

    Telescope   & Instrument         & Distance from   & Approximate peak-to-peak & Amplitude in          & CCD front end   & ADC                           & Reference \\
                &                    & driver pixel to & amplitude of binary      & electrons             &                 &                               & \\
                &                    & target pixel (pixels)      & offset effect (ADU)      &                       &                 &                               & \\
    \hline
    Blanco      & DECam              & -        & Not detected (< \changede{0.05})               & Not detected (\changede{< 0.2})  & Monsoon         & Analog Devices AD7674         & \citet{castilla10} \\
    \hline
    CFHT        & Megacam            & 1        & 0.4                                & 0.6                   &                 & Linear Technology LTC 1604    & \citet{dekat04} \\
    \hline
    Gemini      & GMOS-S E2V         & 2, 3     & 0.7                                & 1.4                   & ARC Gen. II     & Datel ADS-937                 & \changede{\citet{hook04}} \\
                & GMOS-S Hamamatsu   & -        & Not detected (< 0.05)              & Not detected (< 0.08) & ARC Gen. III     &                               & \changede{\citet{gimeno16}} \\
                & GMOS-N E2V         & 2, 3     & 0.7                                & 1.4                   & ARC Gen. II     & Datel ADS-937                 & \changede{\citet{hook04}} \\
                & GMOS-N Hamamatsu   & -        & Not detected (< 0.05)              & Not detected (< 0.09) & ARC Gen. III     &                               & \changede{\citet{gimeno16}} \\
    \hline
    HST         & WFC3 UVIS          & 1        & 0.05-0.15$^a$                         & 0.08-0.23$^a$            &                 &                               & \\
                & STIS (post SM4)    & 1        & 0.5-4.5$^a$                           & 0.5-4.5$^a$              &                 &                               & \\
                & ACS                & 1, 2, 3, 4   & 0.4-1.0$^a$                           & 0.4-1.0$^a$              &                 &                               & \\
    \hline
    Keck        & DEIMOS             & 2        & 2.6                                & 3.2                   & ARC Gen. II     & Datel ADS-937                 & \citet{wright03} \\
                & HIRES              & 2        & 0.3                                & 0.6                   & ARC Gen. I      & Datel ADS-937                 & \citet{kibrick93} \\
                & LRIS B             & 2        & 0.15                               & 0.24                  & ARC Gen. I      & Datel ADS-937                 & \citet{mccarthy98} \\
                & LRIS R (upgraded)  & 2        & 0.15                               & 0.15                  & ARC Gen. II     & Datel ADS-937                 & \citet{rockosi10} \\
    \hline
    SDSS        & -                  & -        & Not detected (< 0.05)              & Not detected (< 0.23) &                 & Crystal Semiconductor CS5101A & \citet{gunn98} \\
    \hline
    Subaru      & Suprime-Cam        & -        & Not detected (< 0.05)              & Not detected (< 0.15) & MFront          & Analogic ADC423               & \citet{miyazaki02} \\
                & Hyper Suprime-Cam  & 1        & 0.5$^b$                       & 1.6$^b$          & MFront2         & Analog Devices AD7686C        & \citet{nakaya12} \\
                & FOCAS              & 2        & 0.1$^b$                       & 0.2$^b$          & MFront          & Analogic ADC423               & \citet{kashikawa02} \\
    \hline
    UH88        & SNIFS blue channel & 2, 3     & 2.4                                & \changede{1.8}        & ARC-41 Gen. II  & Datel ADS-937                 & \citet{aldering02} \\
                & SNIFS red channel  & 2, 3     & 1.5                                & \changede{1.1}        & ARC-41 Gen. II  & Datel ADS-937                 & \citet{aldering02} \\
    \hline
    VLT         & FORS 1             & 1        & 0.1                                & 0.22                  & FIERA           & Analogic ADC4320A$^c$             & \citet{beletic98} \\
                & FORS 2             & 1        & 0.1                                & 0.13                  & FIERA           & Analogic ADC4320A$^c$             & \citet{beletic98} \\
                & MUSE               & -        & Not detected (< 0.05)              & Not detected (< 0.06) & NGC             & Analog Devices AD7677$^c$ & \citet{reiss12} \\

    \hline
\end{tabular}
\begin{tablenotes}
    \item $^a$ For the labeled HST instruments, there is a strong trend in the
        mean of the residuals with the least-significant bit
        (odd-even) along with
        large offsets when higher bits change. Intermediate bits do not
        appear to have any effect.
    \item $^b$ In Subaru Hyper Suprime-Cam and FOCAS images we find a
        trend in the mean of the residuals with the least-significant bit (odd-even). The
        higher bits do not appear to have a direct impact on the mean of the residuals.
        \changede{\item $^c$ We thank J. Reyes (ESO) for details of the ADCs used in the VLT instruments. (personal communication, August 2017)}
\end{tablenotes}
\end{threeparttable}
\end{table}

For all of the detections in Table~\ref{tab:offset_size} \changedj{from
    instruments on the CFHT, Gemini, Keck, UH88 and VLT telescopes} (11 of the
22 instruments investigated), we see an offset in the mean of the residuals
that is roughly proportional to the number of "1" bits in the binary
representation of the driver pixel, similar to what is seen for SNIFS. The
amplitude of the effect varies significantly between instruments, from a
peak-to-peak amplitude of approximately 0.1~ADU for FORS 2 on VLT to a
peak-to-peak amplitude of approximately 2.6~ADU for DEIMOS on Keck
\changedj{(see Figure~\ref{fig:other_binary_offset}a)}. We also find
\changedh{that} the number of pixels following the driver pixel at which the
binary offset effect appears varies from 1 to 3 pixels for the different
instruments. All of the instruments tested using ARC Generation I or II
controllers \citep{leach98} show this kind of binary offset effect, although
the size of the effect varies from instrument to instrument. The newer ARC
Generation III controllers used in GMOS-S and GMOS-N do not show any evidence
of the binary offset effect.

In the images from Hyper Suprime-Cam (HSC) on Subaru\changedj{,} we find offsets in the
mean of the residuals, but they do not appear to be linearly related to the number of "1" bits
in the binary encoding of the driver pixel value (see
Figure~\ref{fig:other_binary_offset}\changedj{b}). \changedg{
The mean of the residuals shows peak-to-peak offsets of 0.5~ADU between
adjacent driver pixel values with a median measurement uncertainty of 0.006~ADU,
so the detection of an offset related to the \changedj{driver} pixel value is highly significant.
There appears to be a
correlation of the least-significant bit (i.e. whether the driver pixel value is
even or odd) with the mean of the residuals. Unlike \changedh{the previously-described} instruments,
large changes in the binary representation do not appear to correlate with
large offsets in the mean of the residuals. The introduced offsets do still have a
highly nonlinear and nonmonotonic dependence on the \changedj{driver} pixel values, although
additional work \changedh{is needed} to understand these offsets.}
For images taken with the FOCAS instrument on Subaru we see a similar effect,
although the amplitude is smaller.

Images from all of the HST instruments that were examined (WFC3 UVIS, STIS and
ACS) show evidence of an effect related to the binary encoding of driver
pixels but with somewhat different behavior. An example of the effect in ACS is
shown in Figure~\ref{fig:other_binary_offset}\changedj{c}. The least-significant bit (odd-even)
has a large effect on the mean of the residuals, but transitions of the next several bits
do not appear to have a significant impact. However, when the sixth bit or
higher is changed, a step is introduced into the mean of the residuals. For the ACS images
shown in Figure~\ref{fig:other_binary_offset}\changedj{c}, we see an offset of 0.4~ADU at
the transition from 2463~ADU to 2464~ADU (\changede{1001~1001~1111 to
    1001~1010~0000}).

The STIS instrument on HST displays a similar behavior to ACS. An example of
the binary offset effect on STIS is shown in
\changedh{Figure~\ref{fig:other_binary_offset}}\changedj{d}.
For STIS, when the fifth bit or higher is
changed, we notice a large offset, but there is little effect for the less
significant bits. At the transition from 1343~ADU to 1344~ADU
(\changede{101 0011 1111 to
    101 0100 0000}) there is a 4.5~ADU offset in the mean of the residuals. We find that the size
of the offsets when the upper bits are changed varies dramatically, from 
0.5~ADU up to the offset of 4.5~ADU previously mentioned.

For several instruments, the binary offset effect introduces artifacts that are
large compared to the size of the signals being observed. On SNIFS, the blue
step \changedj{was} identified and characterized before \changedj{it was}
understood as being \changedj{a result} of the binary offset effect. It is
likely that the consequences of the binary offset effect have been identified
in other instruments without the root cause being fully understood.

\section{Modeling and Correcting the Offset for Active Instruments}
\label{sec:model}

The binary offset effect is present in a large fraction of \changedh{existing CCD} data, as
illustrated by Table~\ref{tab:offset_size}, and \changedg{correcting for the effect
will both improve the quality of existing data and allow previously unusable data
to be recovered.} As a working example of such a correction, we have derived a model that can calculate the
introduced offsets in SNIFS data.

As illustrated in Figure~\ref{fig:residuals_vs_binary}, the amplitude of the
effect is primarily proportional to the number of "1" bits in the binary
encoding of the driver pixel. However, this figure illustrates that the effect
is not simply a linear function of the driver pixel value, as the zeropoint of
the \changedj{linear relation} appears to shift when the driver pixel values
drop below 1280~ADU. These details of the effect are second order, but must be
taken into account to generate an accurate model. We have developed a
9-parameter model which can \changedj{capture} the behavior of the binary
offset effect in SNIFS data. We trained this model on a set of bias images
covering the full history of SNIFS operations \changedj{(2004-2017)}. The
details of this model can be found in \changedh{the Appendix}.

We find that \changedh{this} single set of model parameters is able to describe
the behavior of an amplifier over the entire history of the SNIFS instrument.
\changedh{That is;} the model parameters do not vary over time, and a single
set of parameters per amplifier is sufficient to cover the full range of
observed temperatures, bias levels and signal levels. The behavior of the
binary offset effect does vary significantly between amplifiers, and we
therefore use a unique set of model parameters for each amplifier.
\changedk{This is consistent with the idea that the digitization is the root cause,
since each amplifier has its own video board.} With this
model, we are able to predict the amplitude of the offsets introduced by the
binary offset effect to within 0.11-0.16~ADU depending on the amplifier. We can
predict the amplitudes of the offsets for every pixel in an image, and we can
use these predictions to build a correction image which can be
\changedj{subtracted from} the data to remove the binary offset effect.
Applying this procedure to a new image only requires the raw pixel counts of
the image and a set of \changedk{previously-derived} model parameters for that amplifier.
No additional fitting is required.

\begin{figure}
\centering
\includegraphics[scale=0.75]{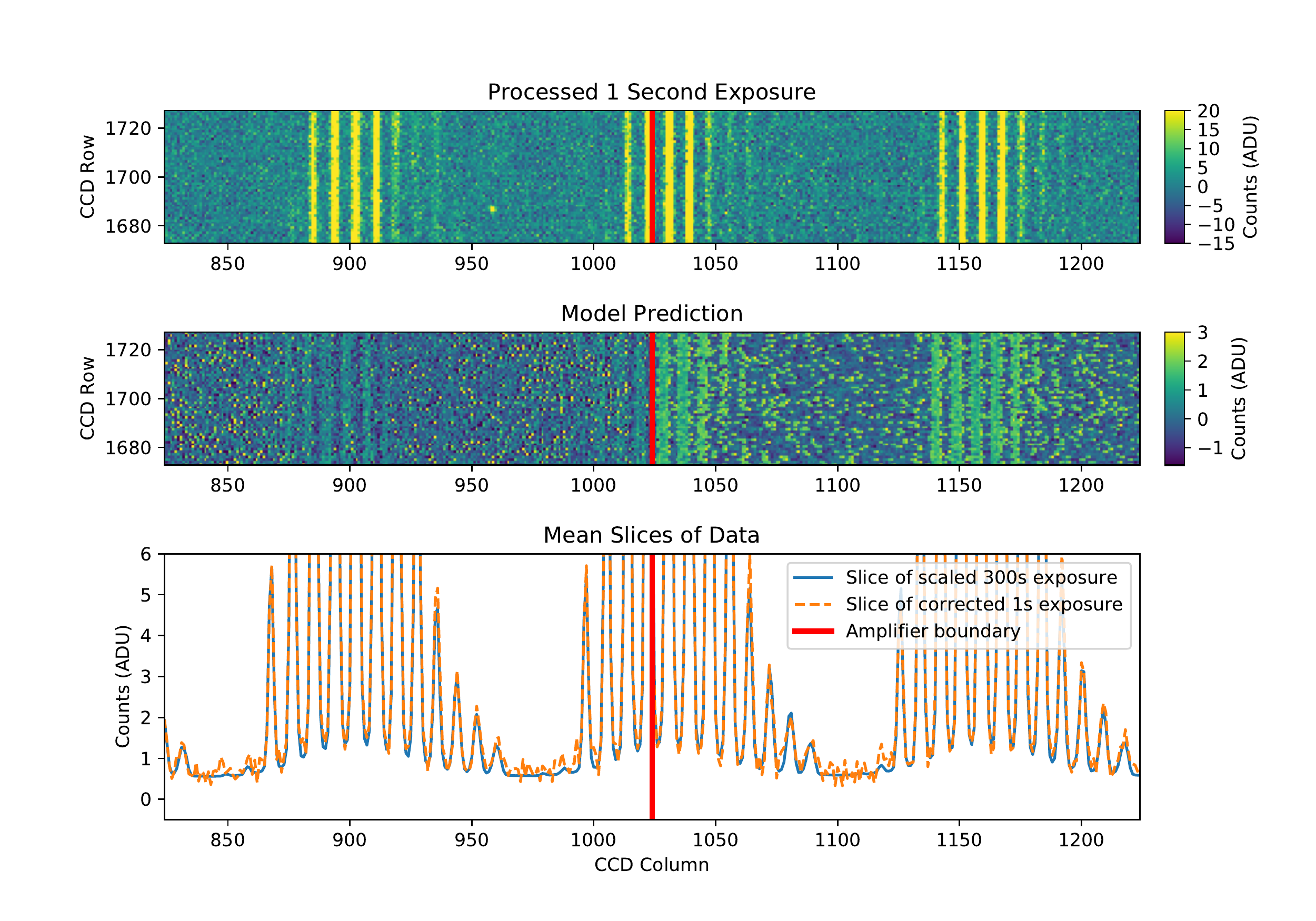}

\caption{
    \changedj{
        \changedk{Results of the binary offset effect model for the SNIFS}
        image shown in Figure~\ref{fig:uncorrected2d}. Top panel:
        \changedk{original} 1 second exposure.
        Middle panel: model prediction of the signal introduced by the binary
        offset effect. Bottom panel: slices through the previous images showing
        the mean values of the pixels in each CCD column with CCD \changedk{row}
        values
        between 1400 and 2000. For the right amplifier, the model predicts that
        the binary offset effect introduces an offset of $\sim$2~ADU \changedk{($\sim$1.3 electrons)} at the
        specific locations on the CCD where spectral traces are found relative
        to the background locations. When these corrections are applied to the
        data, the count deficits seen in Figure~\ref{fig:uncorrected2d} for the
        slice of the 1~second exposure on the right amplifier are eliminated.
    }
}

\label{fig:offset_correction}
\end{figure}

\begin{figure}
\centering
\includegraphics[scale=0.8]{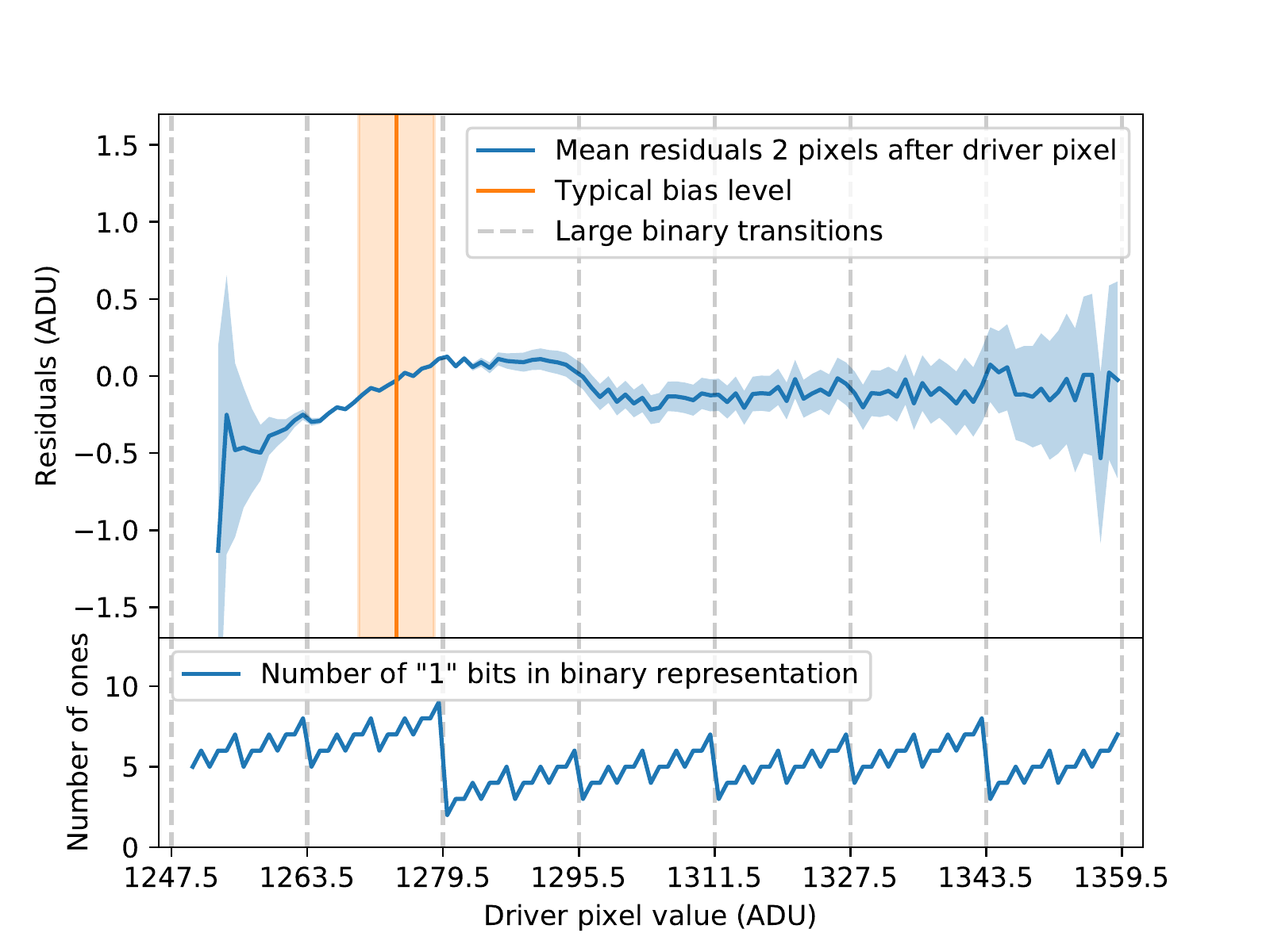}

\caption{Example of data \changedk{that previously exhibited} the binary offset effect on the
    SNIFS blue channel right amplifier, after correcting for the effect. See
    Figure~\ref{fig:binary_offset_effect} for the same data before correction
    and an explanation of how this figure was produced. The residuals after
    correction no longer have a strong dependence on the binary encoding. There
    is a slight linear slope in the residuals for driver pixel values between
    $\sim$1250-1280 ADU\changedk{; this} slope arises due to correlations introduced by
    high frequency pickup, as discussed in \changedj{the Appendix}\changedk{,} and is not
    related to the binary offset effect.}

\label{fig:binary_offset_effect_corrected}
\end{figure}

As a test of the model, we applied the derived correction to the set of 1 and
300 second dome flat exposures described in Section~\ref{sec:evidence}.
\changedj{The results of this procedure can be seen in
    Figure~\ref{fig:offset_correction}. The dome flat images were not included the
    dataset on which the model was trained, so this is an out-of-sample test
    of the model. The deficits seen on the right amplifier in the slice plot of
    Figure~\ref{fig:uncorrected2d} are no longer present after correction.}
We applied the same algorithm used to produce
Figure~\ref{fig:binary_offset_effect} to the corrected data. The results of
this procedure are shown in Figure~\ref{fig:binary_offset_effect_corrected}. As
seen from this figure, \changedj{the offsets introduced by the binary offset
    effect} are effectively removed from the data when the correction is
applied. The remaining linear slope in the residuals is a consequence of the
high frequency pickup described in the Appendix, and is not a feature of the
binary offset effect.

\section{Impact of the Binary Offset Effect on Scientific Results}
\label{sec:impact}

\subsection{Impact of the Binary Offset Effect on SNIFS/SNfactory Data}

\changedj{
    The binary offset effect introduces a highly non-linear signal into
    CCD data that affects science results in several ways. As discussed in
    Section~\ref{sec:implications_snifs}, for the SNIFS instrument, the binary
    offset effect can introduce an offset to the data that only appears at
    locations on the CCD with a significant amount of flux. This can be
    seen directly in the model in Figure~\ref{fig:offset_correction}: on the
    right amplifier, there is a visible offset of $\sim$2~ADU \changedk{($\sim$1.3 electrons)}
    at the specific
    locations on the CCD where spectral traces are found relative to the
    background regions. The model predicts that the binary offset effect has
    little impact on the left amplifier due to the bias level being above the
    large binary transition at 1280.
}

\begin{figure}
\centering
\includegraphics[scale=0.8]{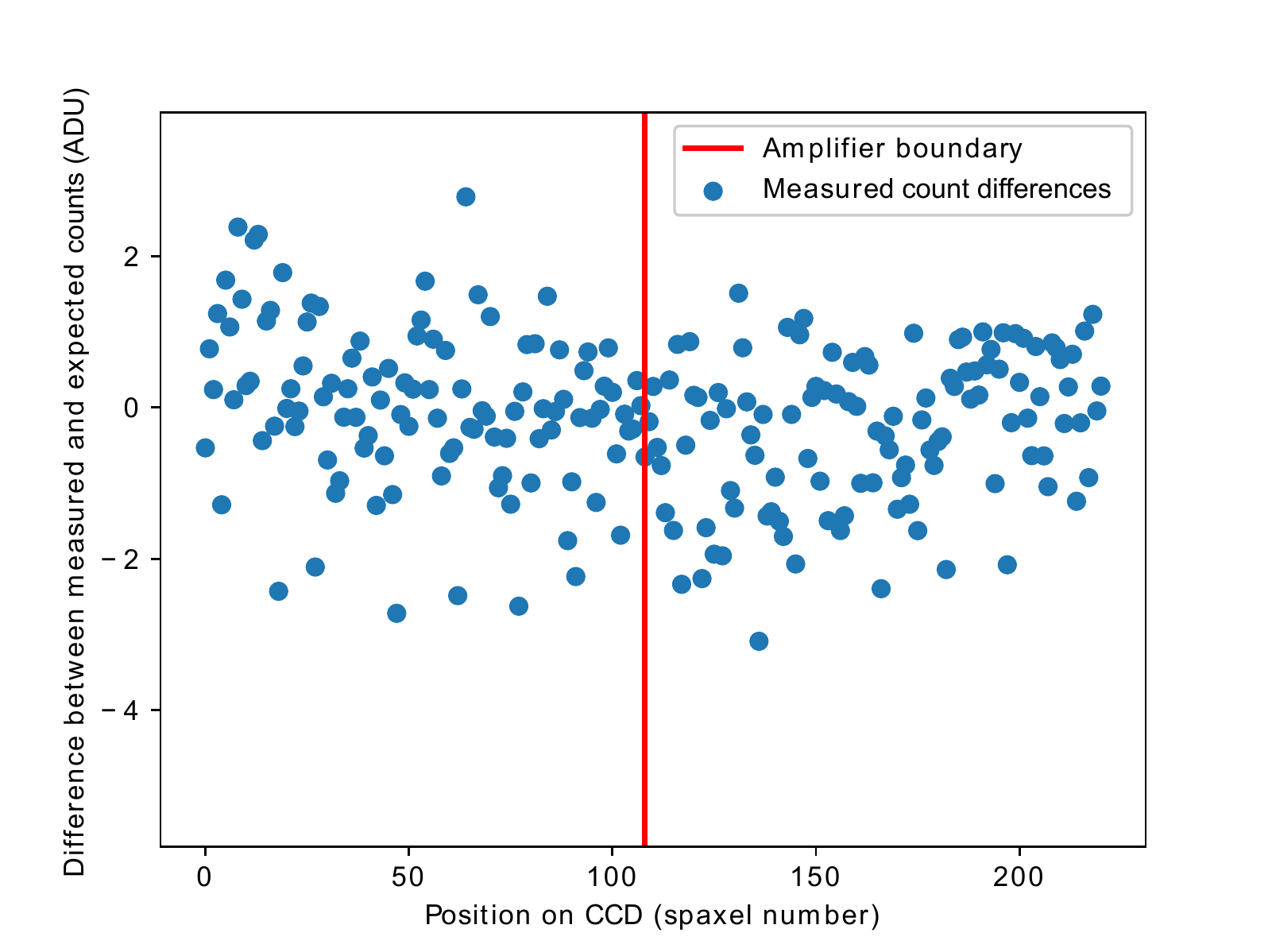}

\caption{Corrected exposure-time normalized difference between extracted 1 second
    and 300 second dome flats from the SNIFS \changede{blue} channel. See
    Figure~\ref{fig:rawfluxratio} for details and comparison. We find that the offsets
    between the two amplifiers are consistent to within 0.2~ADU (0.13 electrons) after
    correction.}

\label{fig:corr_flux_ratio}
\end{figure}

\changedj{
    We subtract the derived model from one of the 1~second dome flat exposures,
    and we extract the spectra in the corrected image \changedk{using} the SNfactory
    pipeline. The result of this procedure is shown in
    Figure~\ref{fig:corr_flux_ratio}, which can be compared to
    Figure~\ref{fig:rawfluxratio} on the same data before corrections. The
    previous offset of 2.2~ADU has been removed and we constrain the corrected
    offset between the amplifiers to be less than 0.2~ADU \changedk{(0.13 electrons)},
    confirming that the
    binary offset effect was the cause of the "blue step" in this data and that
    our model is capable of removing it.
}

We \changedj{apply} the binary offset correction to a series of SNfactory
observations of Type~Ia supernovae that are representative of the range of
measured flux levels on the CCD for scientific observations. We find that the
blue channel is the most strongly affected by the binary offset effect. For the
full dataset, the NMAD \changedj{(normalized median absolute deviation)} of the
applied corrections ranges is 0.04~mag at UV wavelengths and 0.02~mag for blue
(B-band) wavelengths, while if we look at the faintest 20th percentile of data
ordered by flux level, we find that the corrections in UV have a much larger
NMAD of 0.51~mag and those in the blue have an NMAD of 0.11~mag. Corrections on
the red channel have an NMAD of less than 0.008~mag over the full dataset, and
an NMAD of less than 0.05~mag for the faintest 20 percent of observations. The
binary offset correction is therefore relatively small for the majority of the
dataset, but it has a disproportionately high impact on fainter
\changedk{regions of our spectra on the blue channel}.

The affected spectra in SNfactory data are typically at late phases of the
supernova lightcurves where the intrinsic supernova flux is low. In practice,
however, the SNfactory collaboration had previously developed methods to
identify spectra affected by the binary offset effect (before recognizing the
cause), primarily by looking for anomalous behavior in the UV, where it stands
out. These spectra were not included in previous published analyses. Although
the corrections for the binary offset thus should not affect any previous
results, we do expect that it will allow future analyses to include many
spectra that were previously rejected, especially spectra at late phases.

\subsection{Impact of the Binary Offset Effect on General Astronomical Data}
    
There are several science scenarios where the binary offset effect can have
large impacts on science data. The conditions that lead to the largest
potential impact are:

\begin{itemize}
    \item Low background noise.
    \item Low signal levels.
    \item A background level just below a large binary transition.
\end{itemize}

Low-flux observations with \changedj{fiber-fed, lenslet, or similar}
spectrographs are very susceptible to this effect. In these cases, most of the
CCD is not in the path of any light, so the background in those regions is
entirely dominated by read noise. For modern instruments, this means that the
background noise will be very low, on the order of a few ADU or less. When
measuring the background level on the CCD with conventional methods, one will
effectively sample the binary offset curve at the given bias level instead of
measuring what the background truly is under the science data due to the small
range of pixel values covered in the region used to measure the background
level. Observations with ground-based slit spectrographs may be highly
susceptible by the binary offset effect, although their susceptibility depends
on how they are used. Observations of faint emission lines with low sky
background or target continuum will be affected for the same reasons as
fiber-fed and IFU spectrographs. As the binary offset effect introduces an
offset that is shifted by 1-3 pixel from the data that it is applied to, the
recovered wavelengths of emission lines may be biased. Higher resolution
spectrographs are likely to be more strongly affected because of the reduction
in sky background per pixel leading to low background noise levels on the CCD.

\changedk{Science applications using ground-based CCD imaging} are not as
susceptible as spectrographs to the
binary offset effect. Measurements of faint targets with these detectors are
often limited by the sky background, so the potential size of the offset is
suppressed. Imaging applications that have low backgrounds and low signals (for
example U-band observations, narrow band imaging, \changedj{imaging of standard
    stars, or very short} exposures) are more likely to be affected. An
additional issue for imaging is that the local background offset will have the
same shape as the science signal, but will be shifted by 1-3 pixels depending
on the instrument. This can affect measurements of the shape of a galaxy and
may be a concern for some weak lensing analyses. The binary offset effect will
also introduce some correlations between nearby pixels on the CCD. 

\changedk{Science applications using detectors} in space-based missions are highly
susceptible to the binary offset effect, regardless of whether they are
being used as an imager or any type of spectrograph. For these instruments, the sky noise is
often low compared to the readout noise so conventional background subtraction
routines will effectively sample from a point on the binary offset curve rather
than averaging over it. The STIS spectrograph on HST is of particular interest.
We find an offset of up to 4.5~ADU for some binary transitions in this
instrument. This offset is per pixel, so it can easily add up to a significant
fraction of the science signal.

\section{Conclusions}

We have discovered an anomalous behavior in the read-out of CCDs: the
introduction of \changedj{spurious counts} into a pixel, with \changedj{an
    amplitude that depends} on the binary encoding of a pixel read out 1-3
pixels previously. One consequence of this effect is that it can introduce a
local background offset that only appears where \changedj{sufficient flux from
    science targets is} present on the CCD. This background offset is not
removed by conventional background subtraction procedures. In SNIFS data, the
effect \changedh{can cause} an offset in the final measured fluxes of up to
2~ADU per pixel. The binary offset effect explains several effects previously
noticed in SNIFS data, notably the "blue step" where there is a flat offset
between data from different amplifiers, and a fixed count deficit in pixels
following cosmic rays. We find evidence of the binary offset effect in 16 of 22
different instruments that were investigated, indicating that the effect is
present in a significant amount of \changedh{existing} astronomical data.

In this paper, we built a model that can predict, and thus reliably correct,
the offsets introduced by this effect in affected CCD data\changedj{, and we
    applied this model to SNIFS data}. We find that the model parameters are
stable in time, but that they are unique to \changedf{the readout electronics
    of} each amplifier. With this model, we can predict the amplitudes of the
introduced offsets, and we use these predictions to correct for the binary
offset effect. The model derived in this paper is general, and a variant of it
should be \changedf{applicable} to any instrument. \changedj{Because of the
    relatively large amplitudes of the introduced offsets relative to the noise
    levels}, only a single bias image is typically required to characterize the
effect and fit the model.

The binary offset effect can also be mitigated in \changedf{hardware.}
Reducing opportunities
for stray signals to affect the input of the ADC, e.g. using a differential input, should
eliminate the binary offset effect. We find
that 6 of the 22 instruments that we investigated do not show signs of this
effect, as shown in Table~\ref{tab:offset_size}. These 
can serve as a guide to design new readout electronics that aren't susceptible to
the binary offset effect. The size of the binary
offset effect should be measured and characterized in any new CCD readout
system to determine whether or not it will impact the science results. For systems where
modifying the readout electronics is not practical, the bias level can at least
be set to a level where there are no large binary transitions
\changedj{immediately above said bias level}. This will
minimize the potential size of the offsets although it will not eliminate them.
Finally, the tests outlined in this paper should be performed for any CCD-based
instrument likely to exhibit the binary offset effect, and the correction procedure
that we present in this paper should be incorporated in the data-reduction pipeline 
of all instruments where the binary offset effect is found.

\section{Acknowledgements}



\changede{
We thank the technical staff of the University of Hawaii 2.2 m telescope, and Dan Birchall for observing assistance.
The authors wish to recognize and acknowledge the very significant cultural role and reverence that the summit of Maunakea has always had within the indigenous Hawaiian community.  We are most fortunate to have the opportunity to conduct observations from this mountain.
This work was supported in part by the
Director, Office of Science, Office of High Energy Physics of the U.S. Department of Energy under Contract No.
DE-AC02-05CH11231.
Support in France was provided by CNRS/IN2P3, CNRS/INSU, and PNC; LPNHE acknowledges support
from LABEX ILP, supported by French state funds managed by the ANR within the Investissements d'Avenir programme
under reference ANR-11-IDEX-0004-02.
%
%
%
%
Some results were obtained using resources and support from the National Energy Research Scientific Computing Center,
supported by the Director, Office of Science, Office of Advanced Scientific Computing Research of the
U.S. Department of Energy under Contract No. DE-AC02- 05CH11231.
We thank the Gordon \& Betty Moore Foundation for their continuing support.
%
%
We also
thank the High Performance Research and Education Network (HPWREN), supported by National Science Foundation
Grant Nos. 0087344 \& 0426879. 
}


\changede{
Funding for SDSS-III has been provided by the Alfred P. Sloan Foundation, the Participating Institutions, the National Science Foundation, and the U.S. Department of Energy Office of Science. The SDSS-III web site is http://www.sdss3.org/. SDSS-III is managed by the Astrophysical Research Consortium for the Participating Institutions of the SDSS-III Collaboration including the University of Arizona, the Brazilian Participation Group, Brookhaven National Laboratory, Carnegie Mellon University, University of Florida, the French Participation Group, the German Participation Group, Harvard University, the Instituto de Astrofisica de Canarias, the Michigan State/Notre Dame/JINA Participation Group, Johns Hopkins University, Lawrence Berkeley National Laboratory, Max Planck Institute for Astrophysics, Max Planck Institute for Extraterrestrial Physics, New Mexico State University, New York University, Ohio State University, Pennsylvania State University, University of Portsmouth, Princeton University, the Spanish Participation Group, University of Tokyo, University of Utah, Vanderbilt University, University of Virginia, University of Washington, and Yale University.
}


\changede{
This project used public archival data from the Dark Energy Survey (DES). Funding for the DES Projects has been provided by the U.S. Department of Energy, the U.S. National Science Foundation, the Ministry of Science and Education of Spain, the Science and Technology Facilities Council of the United Kingdom, the Higher Education Funding Council for England, the National Center for Supercomputing Applications at the University of Illinois at Urbana-Champaign, the Kavli Institute of Cosmological Physics at the University of Chicago, the Center for Cosmology and Astro-Particle Physics at the Ohio State University, the Mitchell Institute for Fundamental Physics and Astronomy at Texas A\&M University, Financiadora de Estudos e Projetos, Funda\c{c}\~{a}o Carlos Chagas Filho de Amparo \`{a} Pesquisa do Estado do Rio de Janeiro, Conselho Nacional de Desenvolvimento Cient\'{i}fico e Tecnol\'{o}gico and the Minist\'{e}rio da Ci\^{e}ncia, Tecnologia e Inova\c{c}\~{a}o, the Deutsche Forschungsgemeinschaft and the Collaborating Institutions in the Dark Energy Survey. The Collaborating Institutions are Argonne National Laboratory, the University of California at Santa Cruz, the University of Cambridge, Centro de Investigaciones En\'{e}rgeticas, Medioambientales y Tecnol\'{o}gicas-Madrid, the University of Chicago, University College London, the DES-Brazil Consortium, the University of Edinburgh, the Eidgen\"{o}ssische Technische Hochschule (ETH) Z\"{u}rich, Fermi National Accelerator Laboratory, the University of Illinois at Urbana-Champaign, the Institut de Ci\`{e}ncies de l'Espai (IEEC/CSIC), the Institut de F\'{i}sica d'Altes Energies, Lawrence Berkeley National Laboratory, the Ludwig-Maximilians Universit\"{a}t M\"{u}nchen and the associated Excellence Cluster Universe, the University of Michigan, the National Optical Astronomy Observatory, the University of Nottingham, the Ohio State University, the University of Pennsylvania, the University of Portsmouth, SLAC National Accelerator Laboratory, Stanford University, the University of Sussex, and Texas A\&M University. 
Based in part on observations at Cerro Tololo Inter-American Observatory, National Optical Astronomy Observatory (NOAO Prop. I D and PI), which is operated by the Association of Universities for Research in Astronomy (AURA) under a cooperative agreement with the National Science Foundation. 
}


\changede{
\changedf{Based in part} on observations obtained with MegaPrime/MegaCam, a joint project of CFHT and CEA/DAPNIA, at the Canada-France-Hawaii Telescope (CFHT) which is operated by the National Research Council (NRC) of Canada, the Institut National des Sciences de l'Univers of the Centre National de la Recherche Scientifique (CNRS) of France, and the University of Hawaii.
}


\changede{
\changedf{Based in part} on observations obtained at the Gemini Observatory acquired through the Gemini Observatory Archive, which is operated by the Association of Universities for Research in Astronomy, Inc., under a cooperative agreement with the NSF on behalf of the Gemini partnership: the National Science Foundation (United States), the National Research Council (Canada), CONICYT (Chile), Ministerio de Ciencia, Tecnolog\'{i}a e Innovaci\'{o}n Productiva (Argentina), and Minist\'{e}rio da Ci\^{e}ncia, Tecnologia e Inova\c{c}\~{a}o (Brazil).
}


\changede{
\changedf{Based in part} on observations made with the NASA/ESA Hubble Space Telescope, obtained from the Data Archive at the Space Telescope Science Institute, which is operated by the Association of Universities for Research in Astronomy, Inc., under NASA contract NAS 5-26555. These observations are associated with programs 9583, 13677, 14327, and 14820.
}


\changede{
Some of the data presented herein were obtained at the W. M. Keck Observatory, which is operated as a scientific partnership among the California Institute of Technology, the University of California and the National Aeronautics and Space Administration. The Observatory was made possible by the generous financial support of the W. M. Keck Foundation.
This research has made use of the Keck Observatory Archive (KOA), which is operated by the W. M. Keck Observatory and the NASA Exoplanet Science Institute (NExScI), under contract with the National Aeronautics and Space Administration.
}


\changede{
Based in part on data collected at Subaru Telescope, which is operated by the National Astronomical Observatory of Japan.
}


\changede{
\changedf{Based in part on} data obtained from the ESO Science Archive Facility under requests numbered kboone301279, kboone301093, and kboone301092.
}

\bibliographystyle{yahapj}
\bibliography{references}

\begin{thebibliography}{}
\providecommand\natexlab[1]{#1}
\providecommand\JournalTitle[1]{#1}

\bibitem[{{Aldering} {et~al.}(2002){Aldering}, {Adam}, {Antilogus}, {Astier},
  {Bacon}, {Bongard}, {Bonnaud}, {Copin}, {Hardin}, {Henault}, {Howell},
  {Lemonnier}, {Levy}, {Loken}, {Nugent}, {Pain}, {Pecontal}, {Pecontal},
  {Perlmutter}, {Quimby}, {Schahmaneche}, {Smadja}, \&
  {Wood-Vasey}}]{aldering02}
{Aldering}, G., {Adam}, G., {Antilogus}, P., {et~al.} 2002,
  \href{http://dx.doi.org/10.1117/12.458107}{in \procspie, Vol. 4836, Survey
  and Other Telescope Technologies and Discoveries, ed. J.~A. {Tyson} \&
  S.~{Wolff}}, 61

\bibitem[{{Aldering} {et~al.}(2006){Aldering}, {Antilogus}, {Bailey}, {Baltay},
  {Bauer}, {Blanc}, {Bongard}, {Copin}, {Gangler}, {Gilles}, {Kessler},
  {Kocevski}, {Lee}, {Loken}, {Nugent}, {Pain}, {P{\'e}contal}, {Pereira},
  {Perlmutter}, {Rabinowitz}, {Rigaudier}, {Scalzo}, {Smadja}, {Thomas},
  {Wang}, {Weaver}, \& {Nearby Supernova Factory}}]{aldering06}
{Aldering}, G., {Antilogus}, P., {Bailey}, S., {et~al.} 2006,
  \href{http://dx.doi.org/10.1086/507020}{\JournalTitle{\apj}, 650, 510}

\bibitem[{{Antilogus} {et~al.}(2014){Antilogus}, {Astier}, {Doherty},
  {Guyonnet}, \& {Regnault}}]{antilogus04}
{Antilogus}, P., {Astier}, P., {Doherty}, P., {Guyonnet}, A., \& {Regnault}, N.
  2014,
  \href{http://dx.doi.org/10.1088/1748-0221/9/03/C03048}{\JournalTitle{Journal
  of Instrumentation}, 9, C03048}

\bibitem[{{Baggett} {et~al.}(2004){Baggett}, {Hartig}, \& {Cheung}}]{baggett04}
{Baggett}, S., {Hartig}, G., \& {Cheung}, E. 2004, {WFC3 UVIS Crosstalk
  Images}, Tech. rep.

\bibitem[{{Baggett} {et~al.}(2012){Baggett}, {Noeske}, {Anderson}, {MacKenty},
  \& {Petro}}]{baggett12}
{Baggett}, S.~M., {Noeske}, K., {Anderson}, J., {MacKenty}, J.~W., \& {Petro},
  L. 2012, \href{http://dx.doi.org/10.1117/12.926901}{in \procspie, Vol. 8453,
  High Energy, Optical, and Infrared Detectors for Astronomy V}, 845336

\bibitem[{Barbary(2016)}]{barbary16}
Barbary, K. 2016,
  \href{http://dx.doi.org/10.21105/joss.00058}{\JournalTitle{The Journal of
  Open Source Software}, 1}

\bibitem[{{Beletic} {et~al.}(1998){Beletic}, {Gerdes}, \&
  {Duvarney}}]{beletic98}
{Beletic}, J.~W., {Gerdes}, R., \& {Duvarney}, R.~C. 1998,
  \href{http://dx.doi.org/10.1007/978-94-011-5262-4_17}{in Astrophysics and
  Space Science Library, Vol. 228, Optical Detectors for Astronomy, ed.
  J.~{Beletic} \& P.~{Amico}}, 103

\bibitem[{{Bertin} \& {Arnouts}(1996)}]{bertin96}
{Bertin}, E., \& {Arnouts}, S. 1996,
  \href{http://dx.doi.org/10.1051/aas:1996164}{\JournalTitle{\aaps}, 117, 393}

\bibitem[{{Brown} \& {Lupie}(2004)}]{brown04}
{Brown}, T.~M., \& {Lupie}, O. 2004, {Filter Ghosts in the WFC3 UVIS Channel},
  Tech. rep.

\bibitem[{{Caldwell} {et~al.}(2010){Caldwell}, {Kolodziejczak}, {Van Cleve},
  {Jenkins}, {Gazis}, {Argabright}, {Bachtell}, {Dunham}, {Geary}, {Gilliland},
  {Chandrasekaran}, {Li}, {Tenenbaum}, {Wu}, {Borucki}, {Bryson}, {Dotson},
  {Haas}, \& {Koch}}]{caldwell10}
{Caldwell}, D.~A., {Kolodziejczak}, J.~J., {Van Cleve}, J.~E., {et~al.} 2010,
  \href{http://dx.doi.org/10.1088/2041-8205/713/2/L92}{\JournalTitle{\apjl},
  713, L92}

\bibitem[{{Castilla} {et~al.}(2010){Castilla}, {Ballester}, {Cardiel},
  {Chappa}, {de Vicente}, {Holm}, {Huffman}, {Kozlovsky}, {Martinez}, {Olsen},
  {Shaw}, \& {Stuermer}}]{castilla10}
{Castilla}, J., {Ballester}, O., {Cardiel}, L., {et~al.} 2010,
  \href{http://dx.doi.org/10.1117/12.856852}{in \procspie, Vol. 7735,
  Ground-based and Airborne Instrumentation for Astronomy III}, 77352O

\bibitem[{{de Kat} {et~al.}(2004){de Kat}, {Boulade}, {Charlot}, {Aune},
  {Borgeaud}, {Carton}, {Eppel{\'e}}, {Gallais}, {Granelli}, {Gros}, {Rousse},
  {Starzynski}, \& {Vigroux}}]{dekat04}
{de Kat}, J., {Boulade}, O., {Charlot}, Abbon, P.~X., {et~al.} 2004,
  \href{http://dx.doi.org/10.1007/978-1-4020-2527-3_69}{in Astrophysics and
  Space Science Library, Vol. 300, Scientific Detectors for Astronomy, The
  Beginning of a New Era, ed. P.~{Amico}, J.~W. {Beletic}, \& J.~E.
  {Belectic}}, 517

\bibitem[{{Dixon} {et~al.}(2016){Dixon}, {Aldering}, {Domagalski}, {Boone},
  {Fagrelius}, {Hayden}, {Perlmutter}, {Saunders}, \& {Sofiatti}}]{dixon16}
{Dixon}, S., {Aldering}, G.~S., {Domagalski}, R., {et~al.} 2016, in American
  Astronomical Society Meeting Abstracts, Vol. 227, American Astronomical
  Society Meeting Abstracts, 146.15

\bibitem[{{Gimeno} {et~al.}(2016){Gimeno}, {Roth}, {Chiboucas}, {Hibon},
  {Boucher}, {White}, {Rippa}, {Labrie}, {Turner}, {Hanna}, {Lazo},
  {P{\'e}rez}, {Rogers}, {Rojas}, {Placco}, \& {Murowinski}}]{gimeno16}
{Gimeno}, G., {Roth}, K., {Chiboucas}, K., {et~al.} 2016,
  \href{http://dx.doi.org/10.1117/12.2233883}{in \procspie, Vol. 9908,
  Ground-based and Airborne Instrumentation for Astronomy VI}, 99082S

\bibitem[{{Gunn} {et~al.}(1998){Gunn}, {Carr}, {Rockosi}, {Sekiguchi}, {Berry},
  {Elms}, {de Haas}, {Ivezi{\'c}}, {Knapp}, {Lupton}, {Pauls}, {Simcoe},
  {Hirsch}, {Sanford}, {Wang}, {York}, {Harris}, {Annis}, {Bartozek},
  {Boroski}, {Bakken}, {Haldeman}, {Kent}, {Holm}, {Holmgren}, {Petravick},
  {Prosapio}, {Rechenmacher}, {Doi}, {Fukugita}, {Shimasaku}, {Okada}, {Hull},
  {Siegmund}, {Mannery}, {Blouke}, {Heidtman}, {Schneider}, {Lucinio}, \&
  {Brinkman}}]{gunn98}
{Gunn}, J.~E., {Carr}, M., {Rockosi}, C., {et~al.} 1998,
  \href{http://dx.doi.org/10.1086/300645}{\JournalTitle{\aj}, 116, 3040}

\bibitem[{{Hook} {et~al.}(2004){Hook}, {J{\o}rgensen}, {Allington-Smith},
  {Davies}, {Metcalfe}, {Murowinski}, \& {Crampton}}]{hook04}
{Hook}, I.~M., {J{\o}rgensen}, I., {Allington-Smith}, J.~R., {et~al.} 2004,
  \href{http://dx.doi.org/10.1086/383624}{\JournalTitle{\pasp}, 116, 425}

\bibitem[{{Janesick}(2001)}]{janesick01}
{Janesick}, J.~R. 2001, {Scientific charge-coupled devices}

\bibitem[{{Jansen} {et~al.}(2003){Jansen}, {Collins}, \&
  {Windhorst}}]{jansen03}
{Jansen}, R.~A., {Collins}, N.~R., \& {Windhorst}, R.~A. 2003, in HST
  Calibration Workshop : Hubble after the Installation of the ACS and the
  NICMOS Cooling System, ed. S.~{Arribas}, A.~{Koekemoer}, \& B.~{Whitmore},
  193

\bibitem[{{Kashikawa} {et~al.}(2002){Kashikawa}, {Aoki}, {Asai}, {Ebizuka},
  {Inata}, {Iye}, {Kawabata}, {Kosugi}, {Ohyama}, {Okita}, {Ozawa}, {Saito},
  {Sasaki}, {Sekiguchi}, {Shimizu}, {Taguchi}, {Takata}, {Yadoumaru}, \&
  {Yoshida}}]{kashikawa02}
{Kashikawa}, N., {Aoki}, K., {Asai}, R., {et~al.} 2002,
  \href{http://dx.doi.org/10.1093/pasj/54.6.819}{\JournalTitle{\pasj}, 54, 819}

\bibitem[{{Kibrick} {et~al.}(1993){Kibrick}, {Stover}, \& {Conrad}}]{kibrick93}
{Kibrick}, R.~I., {Stover}, R.~J., \& {Conrad}, A.~R. 1993, in Astronomical
  Society of the Pacific Conference Series, Vol.~52, Astronomical Data Analysis
  Software and Systems II, ed. R.~J. {Hanisch}, R.~J.~V. {Brissenden}, \&
  J.~{Barnes}, 277

\bibitem[{Kluyver {et~al.}(2016)Kluyver, Ragan-Kelley, P{\'e}rez, Granger,
  Bussonnier, Frederic, Kelley, Hamrick, Grout, Corlay, {et~al.}}]{kluyver16}
Kluyver, T., Ragan-Kelley, B., P{\'e}rez, F., {et~al.} 2016, in Positioning and
  Power in Academic Publishing: Players, Agents and Agendas: Proceedings of the
  20th International Conference on Electronic Publishing, IOS Press, 87

\bibitem[{{Lantz} {et~al.}(2004){Lantz}, {Aldering}, {Antilogus}, {Bonnaud},
  {Capoani}, {Castera}, {Copin}, {Dubet}, {Gangler}, {Henault}, {Lemonnier},
  {Pain}, {Pecontal}, {Pecontal}, \& {Smadja}}]{lantz04}
{Lantz}, B., {Aldering}, G., {Antilogus}, P., {et~al.} 2004,
  \href{http://dx.doi.org/10.1117/12.512493}{in \procspie, Vol. 5249, Optical
  Design and Engineering, ed. L.~{Mazuray}, P.~J. {Rogers}, \& R.~{Wartmann}},
  146

\bibitem[{{Leach} {et~al.}(1998){Leach}, {Beale}, \& {Eriksen}}]{leach98}
{Leach}, R.~W., {Beale}, F.~L., \& {Eriksen}, J.~E. 1998,
  \href{http://dx.doi.org/10.1117/12.316783}{in \procspie, Vol. 3355, Optical
  Astronomical Instrumentation, ed. S.~{D'Odorico}}, 512

\bibitem[{{McCarthy} {et~al.}(1998){McCarthy}, {Cohen}, {Butcher}, {Cromer},
  {Croner}, {Douglas}, {Goeden}, {Grewal}, {Lu}, {Petrie}, {Weng}, {Weber},
  {Koch}, \& {Rodgers}}]{mccarthy98}
{McCarthy}, J.~K., {Cohen}, J.~G., {Butcher}, B., {et~al.} 1998,
  \href{http://dx.doi.org/10.1117/12.316831}{in \procspie, Vol. 3355, Optical
  Astronomical Instrumentation, ed. S.~{D'Odorico}}, 81

\bibitem[{{Miyazaki} {et~al.}(2002){Miyazaki}, {Komiyama}, {Sekiguchi},
  {Okamura}, {Doi}, {Furusawa}, {Hamabe}, {Imi}, {Kimura}, {Nakata}, {Okada},
  {Ouchi}, {Shimasaku}, {Yagi}, \& {Yasuda}}]{miyazaki02}
{Miyazaki}, S., {Komiyama}, Y., {Sekiguchi}, M., {et~al.} 2002,
  \href{http://dx.doi.org/10.1093/pasj/54.6.833}{\JournalTitle{\pasj}, 54, 833}

\bibitem[{{Nakaya} {et~al.}(2012){Nakaya}, {Miyatake}, {Uchida}, {Fujimori},
  {Mineo}, {Aihara}, {Furusawa}, {Kamata}, {Karoji}, {Kawanomoto}, {Komiyama},
  {Miyazaki}, {Morokuma}, {Obuchi}, {Okura}, {Tanaka}, {Tanaka}, {Uraguchi}, \&
  {Utsumi}}]{nakaya12}
{Nakaya}, H., {Miyatake}, H., {Uchida}, T., {et~al.} 2012,
  \href{http://dx.doi.org/10.1117/12.925764}{in \procspie, Vol. 8453, High
  Energy, Optical, and Infrared Detectors for Astronomy V}, 84532R

\bibitem[{{Reiss} {et~al.}(2012){Reiss}, {Deiries}, {Lizon}, \&
  {Rupprecht}}]{reiss12}
{Reiss}, R., {Deiries}, S., {Lizon}, J.-L., \& {Rupprecht}, G. 2012,
  \href{http://dx.doi.org/10.1117/12.925388}{in \procspie, Vol. 8446,
  Ground-based and Airborne Instrumentation for Astronomy IV}, 84462P

\bibitem[{{Robberto} \& {Hilbert}(2005)}]{robberto05}
{Robberto}, M., \& {Hilbert}, B. 2005, {The behaviour of the WFC3 UVIS and IR
  Analog-to-Digital Converters}, Tech. rep.

\bibitem[{{Rockosi} {et~al.}(2010){Rockosi}, {Stover}, {Kibrick}, {Lockwood},
  {Peck}, {Cowley}, {Bolte}, {Adkins}, {Alcott}, {Allen}, {Brown}, {Cabak},
  {Deich}, {Hilyard}, {Kassis}, {Lanclos}, {Lewis}, {Pfister}, {Phillips},
  {Robinson}, {Saylor}, {Thompson}, {Ward}, {Wei}, \& {Wright}}]{rockosi10}
{Rockosi}, C., {Stover}, R., {Kibrick}, R., {et~al.} 2010,
  \href{http://dx.doi.org/10.1117/12.856818}{in \procspie, Vol. 7735,
  Ground-based and Airborne Instrumentation for Astronomy III}, 77350R

\bibitem[{{Scalzo} {et~al.}(2010){Scalzo}, {Aldering}, {Antilogus}, {Aragon},
  {Bailey}, {Baltay}, {Bongard}, {Buton}, {Childress}, {Chotard}, {Copin},
  {Fakhouri}, {Gal-Yam}, {Gangler}, {Hoyer}, {Kasliwal}, {Loken}, {Nugent},
  {Pain}, {P{\'e}contal}, {Pereira}, {Perlmutter}, {Rabinowitz}, {Rau},
  {Rigaudier}, {Runge}, {Smadja}, {Tao}, {Thomas}, {Weaver}, \&
  {Wu}}]{scalzo10}
{Scalzo}, R.~A., {Aldering}, G., {Antilogus}, P., {et~al.} 2010,
  \href{http://dx.doi.org/10.1088/0004-637X/713/2/1073}{\JournalTitle{\apj},
  713, 1073}

\bibitem[{{Stubbs}(2014)}]{stubbs14}
{Stubbs}, C.~W. 2014,
  \href{http://dx.doi.org/10.1088/1748-0221/9/03/C03032}{\JournalTitle{Journal
  of Instrumentation}, 9, C03032}

\bibitem[{{Wright} {et~al.}(2003){Wright}, {Kibrick}, {Alcott}, {Gilmore},
  {Pfister}, \& {Cowley}}]{wright03}
{Wright}, C.~A., {Kibrick}, R.~I., {Alcott}, B., {et~al.} 2003,
  \href{http://dx.doi.org/10.1117/12.461881}{in \procspie, Vol. 4841,
  Instrument Design and Performance for Optical/Infrared Ground-based
  Telescopes, ed. M.~{Iye} \& A.~F.~M. {Moorwood}}, 214

\end{thebibliography}

\appendix
\changedg{
\section{Modeling the Binary Offset Effect in SNIFS Data}
}

\label{sec:appendix_model}

In SNIFS data, we notice that an offset is introduced into the affected pixel
related to the binary encodings of the driver pixels read out 2 and 3 pixels
earlier. In the following discussion, we refer to a pixel that is affected by
the binary offset as the "target". Each pixel on the CCD serves as a driver
pixel \changedj{for several} other pixels. We subtract a model of the counts on
the CCD from the measured data to obtain a residual image, and we compare the
value of the residual image for each pixel to the raw ADU measured on the
corresponding driver pixels. We refer to driver pixels that were read out N
pixels before the target pixel as T-N. \changedj{There is also some crosstalk
    between the amplifiers: the value of the pixel that was read out 3 pixels
    earlier in time on the other amplifier also affects the introduced offsets.
    We note that the timing of the readout of the other amplifier is what
    matters rather than the physical location on the CCD.} We refer to driver
pixels that were read out N pixels before the target pixel on the other
amplifier as O-N.

We find that the binary codes of the driver pixels interact with each other in
a complex way to produce the offset that is applied to the target pixel. There
is a baseline linear trend of the introduced offsets with the number of "1"
bits in pixels T-2 and T-3. However, there is also some interaction between
these pixels: a bit in pixel T-2 behaves differently if that same bit was on in
pixel T-3 compared to if it was off. The largest offsets occur when a bit is
"1" in both T-2 and T-3. For pixel T-2, we find that the more significant bits
have a larger effect on the introduced offset than the less significant bits.
For pixel T-3, we find evidence of interactions across amplifiers: turning a
bit from "0" to "1" in T-3 introduces an offset that depends on how many bits
were already on in T-3 and in O-3. We model this interaction with a second
order polynomial in the number of bits that are "1" in each of T-3 and O-3. We
do not find any evidence of direct interactions across bits (\changedf{e.g.}:
bit 8 \changedj{does not directly affect the behavior of bit 0}).

We incorporate all of these effects into a model capturing the
\changedj{time-invariant} behavior of the electronics, which is then fit to the
data. The model includes \changedj{9 parameters which are used to predict the
    offsets introduced by the binary offset effect} as a simple function of the
three pixel values T-2, T-3, and O-3. \changedj{The terms can be summarized as
    follows: a base effect in the number of bits in T-2 are "1" that depends on
    whether the same bits are "1" in T-3 or not (2 parameters), larger effects
    for more significant bits in T-2 and T-3 (2 parameters), and a
    2-dimensional polynomial in how many bits are "1" in T-3 and how many are
    "1" in O-3 \changedk{in order} to capture the cross-amplifier effects (5 parameters). The
    zeropoint is arbitrarily chosen such that the model predicts an offset of 0
    when all of the driver pixels are 0.} These terms and the effects that they
capture are shown in Table~\ref{tab:variables}.

\begin{table}
  \centering
  \scriptsize

  \caption{Variables used in the model of the binary offset effect in SNIFS
      data. The full model involves linear corrections for each of these terms.
      The coefficients of these linear corrections differ for the different
      amplifiers, but remain stable over time. In this table we use the
      following notation: T-N refers to the driver pixel read N pixels previously
      and O-N refers to the driver pixel read on the other amplifier N pixels previously.
      \changede{The uncertainties shown on the parameters are the standard deviations of
          fits to 10 different datasets.}}

  \label{tab:variables}

  \changede{
\begin{tabular}{|l|l|l|l|l|l|l|}
  \hline
  Explained effect                                      & Id & Variable & \multicolumn{4}{|c|}{Fitted parameters for SNIFS (ADU/unit \changedj{variable})} \\
  \cline{4-7}
                                                        &    &                                                                                & \multicolumn{2}{|c|}{Blue channel}                                        & \multicolumn{2}{|c|}{Red channel} \\
  \cline{4-7}
                                                        &    &                                                                                & Left amplifier                                                            & Right amplifier                      & Left amplifier     & Right amplifier \\
  \hline
  Base effect                                           & 1  & Number of bits that are "1" in both T-2 and T-3.                               & 0.663 $\pm$ 0.012                                                         & 0.342 $\pm$ 0.006                    & -0.503 $\pm$ 0.006 & -0.091 $\pm$ 0.004   \\
  \cline{2-7}
                                                        & 2  & Number of bits that are \changedj{both} "1" in T-2 and "0" in T-3.             & 0.081 $\pm$ 0.004                                                         & 0.043 $\pm$ 0.005                    & -0.247 $\pm$ 0.005 & -0.140 $\pm$ 0.003  \\
  \hline
  Larger effects for                                    & 3  & Sum of the indices of bits that are "1" in both T-2 and T-3.                   & 0.042 $\pm$ 0.003                                                         & 0.034 $\pm$ 0.002                    & 0.010 $\pm$ 0.007  & 0.022 $\pm$ 0.002    \\
  \cline{2-7}
  more significant bits                                 & 4  & Sum of the indices of bits that are \changedj{both} "1" in T-2 and "0" in T-3. & 0.049 $\pm$ 0.003                                                         & 0.015 $\pm$ 0.003                    & 0.006 $\pm$ 0.007  & 0.021 $\pm$ 0.002    \\
  \hline
  \changedj{2-dimensional}                              & 5  & Number of bits that are "1" in T-3.                                            & -0.410 $\pm$ 0.027                                                        & -0.076 $\pm$ 0.037                   & 0.020 $\pm$ 0.030  & 0.107 $\pm$ 0.023   \\
  \cline{2-7}
  \changedj{polynomial capturing}                       & 6  & Number of bits that are "1" in O-3.                                            & 0.065 $\pm$ 0.017                                                         & -0.036 $\pm$ 0.024                   & 0.144 $\pm$ 0.024  & -0.125 $\pm$ 0.025   \\
  \cline{2-7}
  \changedj{interaction across}                         & 7  & Number of bits that are "1" in \changedj{T-3}, squared.                        & 0.035 $\pm$ 0.002                                                         & 0.024 $\pm$ 0.004                    & 0.019 $\pm$ 0.003  & 0.002 $\pm$ 0.002    \\
  \cline{2-7}
  \changedj{amplifiers}                                 & 8  & Number of bits that are "1" in \changedj{O-3}, squared.                        & 0.037 $\pm$ 0.002                                                         & 0.016 $\pm$ 0.003                    & -0.001 $\pm$ 0.002 & 0.019 $\pm$ 0.003    \\
  \cline{2-7}
                                                        & 9  & (Number of bits that are "1" in T-3)                                           & -0.072 $\pm$ 0.005                                                        & -0.039 $\pm$ 0.006                   & -0.018 $\pm$ 0.003 & -0.020 $\pm$ 0.003\\
                                                        &    & $\times$ (Number of bits that are "1" in O-3)                                  &                                                                           &                                      &                    & \\
  \hline
\end{tabular}
}
\end{table}

A challenge in fitting such a model to data is that effects other than the
binary offset effect can introduce correlations between the observed driver and target
pixels. For example, cosmic rays will produce large signals on the detector that
will show up in the residual images. To mitigate this, we mask out cosmic rays
(and any other bright pixels) by identifying any pixels that are over five standard
deviations above the background noise level in the residual images and flagging a region
with a border of 2 pixels around them. Another challenge is that low-order spatial
variations over the image can also \changedf{introduce} local correlations between pixel values. We
perform a background subtraction on the residual images
using \texttt{sep} \citep{barbary16, bertin96} with
$64 \times 64$ pixel boxes to reduce the potential impact of these variations.
A more challenging feature of the data is a pickup signal that appears in many of the SNIFS
images with a frequency of 15-20 kHz (a period of roughly 3 pixels) and an amplitude of
$\sim$ 1~ADU. This pickup signal is very challenging to model due to its low amplitude,
and it effectively introduces a smooth variation in the mean of the target pixel residuals
as a function of the driver pixel value. We note that the binary offset
effect introduces large offsets into the target pixel value when a driver
pixel value increases by \changedf{1~ADU} while most other effects are continuous. We
therefore fit for the effect of a 1~ADU change in a driver pixel on the target pixel rather
than trying to fit for the offset that was added to each target pixel as a function of the
driver pixel values directly.

\changedj{The final fitting procedure is as follows. Given a raw image, we mask
    out pixels with known issues, and \changedk{then} perform a background subtraction to
    obtain a residual image. For every unique set of three driver pixel values
    T-2, T-3, and O-3, we find all target pixels in the raw image with those
    driver pixel values. We calculate the mean value of the residuals for those
    target pixels and we estimate the uncertainty on that mean value. Note that
    the mean value of the target pixel residuals is a sum of the amplitude of
    the binary offset effect and other effects like pickup that introduce
    correlated residuals. After calculating the mean of the target pixel
    residuals for every combination of driver pixel values, we identify sets of
    driver pixel values where one driver pixel value changed by 1~ADU, and we
    calculate the difference in the mean of the target pixel residuals
    associated with that change in driver pixel value and the measurement
    uncertainty on this difference. We fit our model to these differential
    measurements which mitigates the impact of effects like pickup.
}

We fit the 9-parameter model described in Table~\ref{tab:variables} to a
\changedf{subset} of bias images taken roughly evenly spaced in time across the
full history of the SNIFS instrument \changedj{by performing a $\chi^2$
    minimization}. There are 98 images included in this fit for the blue
channel and 99 images for the red channel. Each sample is split into 10
subsamples, and we fit the model parameters on each of these subsamples
individually. We then calculate the mean model parameters from each of these
fits, and we use the standard deviation of the fitted parameters across
subsamples as an estimate of the systematic uncertainty associated with the
model. The fitted model parameters are shown in Table~\ref{tab:variables}.

We find that when a bit is "1" in both pixels T-2 and T-3, an offset of up to
0.663~$\pm$~0.012~ADU is introduced in the target pixel. The fitted model
parameters are not consistent across amplifiers: on the red channel, the left
amplifier has an offset of -0.503~$\pm$~0.006~ADU per bit that is "1" in both
T-2 and T-3 while the right amplifier has an offset of -0.091~$\pm$~0.004~ADU
per "1" bit. The introduced offsets for the interaction across amplifiers can
be up to 0.144~$\pm$~0.024~ADU per "1" bit for the left amplifier on the red
channel, and the fit finds a strong interaction between the bits in pixels T-3
and O-3 on the blue channel.

The model is not a perfect description of the effect: when fitting the model to
the data, we find that the $\chi^2/DoF$ of the fits is between 1.11 and 1.23
\changedj{when using the measurement uncertainties estimated in the previously
    described procedure}. We estimate the remaining dispersion of the data due
to the binary offset effect by adding an uncertainty term in quadrature
\changedj{to the measurement uncertainty in order} to set the $\chi^2/DoF$
\changedk{to} 1. For the blue channel, this requires an additional 0.156~ADU and 0.164~ADU of
dispersion for the left and right amplifiers respectively. For the red channel,
this requires an additional 0.126~ADU and 0.110~ADU of dispersion for the left
and right amplifiers respectively.

We estimate the uncertainty in the model parameters by taking the standard
deviation of the fits to the 10 different subsets. As a conservative estimate
of the systematic uncertainties, we adopt the standard deviations of the fitted
parameters directly rather than attempting to take out the statistical
component by combining datasets. The derived uncertainties are shown in
Table~\ref{tab:variables}. These uncertainties are highly correlated,
especially the terms relating higher orders in the same bit counts (eg:
parameters 5 and 7) and the terms related to larger effects for more
significant bits (eg: parameters 1 and 3). We estimate the impact of the
variation in the model parameters on the derived corrections by
\changedj{repeatedly sampling realizations of model parameters from a Gaussian
    distribution following the full fitted covariance matrix of the model
    parameters. For each realization of model parameters, we calculate the
    implied corrections for a group of images. Across realizations, we find
    that the standard deviation of the implied corrections is less than
    0.01~ADU for images on the blue channel, and less than 0.005~ADU for images
    on the red channel, so the uncertainties in the model parameters do not
    significantly affect the final corrections.} The uncertainty of the
correction is therefore dominated by the unexplained residual dispersion
described previously, and is between 0.11-0.16 ADU depending on the amplifier.
We do not detect any significant variation in the behavior of the binary offset
effect or the model parameters over the history of SNIFS (from 2004 to 2017).

\end{document}